\documentclass[footinbib,a4paper,aps,prl,reprint,twocolumn,preprintnumbers,amsmath,amssymb,10pt, superscriptaddress]{revtex4-2}
\usepackage{verbatim}
\usepackage{booktabs}
\usepackage{graphicx}
\usepackage{color}
\usepackage{tikz}
\usepackage[colorlinks=true,citecolor=blue,linkcolor=blue,urlcolor=blue]{hyperref}
\usepackage[capitalize]{cleveref} 
\usepackage{braket}
\usepackage{lipsum}  
\usepackage[T1]{fontenc}
\usepackage[normalem]{ulem}
\usepackage{upgreek}
\usepackage[percent]{overpic}
\usepackage{graphicx,xcolor}
\usepackage{dcolumn}
\usepackage{bm}
\usepackage[shortlabels]{enumitem}
\graphicspath{{Figs/}}
\makeatletter
\renewcommand{\fnum@figure}{FIG. \thefigure}
\makeatother

\newcommand{\bZ}[1]{\mathbb{Z}}

\begin{document}

\title{Two-dimensional $J_1$-$J_2$ clock model: Enhanced symmetries, emergent orders, and Landau-incompatible transitions}

\author{Pulloor Kuttanikkad Vishnu}
\affiliation{Department of Physics, Indian Institute of Technology Madras, Chennai 600036, India}
\affiliation{Center for Quantum Information, Communication and Computation (CQuICC), Indian Institute of Technology Madras, Chennai 600036, India}

\author{Abhishodh Prakash}
\affiliation{Harish-Chandra Research Institute, a CI of Homi Bhabha National Institute, Prayagraj (Allahabad) - 211019, India}

\author{Rajesh Narayanan}
\affiliation{Department of Physics, Indian Institute of Technology Madras, Chennai 600036, India}
\affiliation{Center for Quantum Information, Communication and Computation (CQuICC), Indian Institute of Technology Madras, Chennai 600036, India}

\author{Titas Chanda}
\email{titas.chanda@physics.iitm.ac.in}
\affiliation{Department of Physics, Indian Institute of Technology Madras, Chennai 600036, India}
\affiliation{Center for Quantum Information, Communication and Computation (CQuICC), Indian Institute of Technology Madras, Chennai 600036, India}

\date{\today}

\begin{abstract}
We present a comprehensive study on the frustrated $J_1$‑$J_2$ classical $q$‑state clock model with even $q>4$ on a two‑dimensional square lattice, revealing a rich ensemble of phases driven by competing interactions. In the unfrustrated regime ($J_1>2J_2$), the model reproduces the standard clock model phenomenology: a low‑temperature $\mathbb{Z}_q$‑broken ferromagnet, an intermediate XY‑like critical quasi‑long‑range–ordered (QLRO) phase with emergent $U(1)$ symmetry, and a high‑temperature paramagnet. For $J_1<2J_2$, frustration stabilizes five distinct regimes: the disordered paramagnet, a stripe‑ordered phase breaking $\mathbb{Z}_q\times\mathbb{Z}_2$ symmetry, two $\mathbb{Z}_2$‑broken nematic phases (one with and one without QLRO), and an exotic stripe phase with emergent discrete $\mathbb{Z}_q$ spin degrees of freedom prohibited in the microscopic Hamiltonian.
Remarkably, this \textit{seemingly forbidden} $\mathbb{Z}_q$ order emerges via a relevant operator in the infrared long-wavelength limit, rather than from an irrelevant perturbation, highlighting a non-standard route to emergence. Using large‑scale corner transfer matrix renormalization group calculations, complemented by classical Monte Carlo simulations, we map the complete phase diagram and identify Berezinskii–Kosterlitz–Thouless, Ising, first‑order, and unconventional Landau‑incompatible transitions between different phases. Finally, we propose an effective field‑theoretic framework that encompasses these emergent orders and their interwoven transitions.
\end{abstract}

\maketitle

\paragraph{Introduction.--}
Modern many-body physics is replete with examples of emergence \cite{Witten2018,Coldea_Science_2010,Schulz_nature} wherein collective behavior of a many-body system, as described via a long-wavelength or coarse-grained description, hosts a higher symmetry in comparison to the bare microscopic theory. 
Field theoretically such emergent symmetries are typically predicated on the presence of an operator that becomes irrelevant (in the renormalization group (RG) sense) in the infrared regime, leading to either a critical point or even an entire critical phase hosting a higher symmetry than the short-wavelength microscopic theory~\cite{Giamarchi2003, Gogolin2004, Shankar2017, Mussardo2020}. Causes c\'el\`ebres of systems hosting such emergent behavior are manyfold and include, for example, the two-dimensional (2D) ferromagnetic $q$-state classical clock model (with $q>4$)~\cite{Jose1977, Elitzur1979, Cardy1980, Tobochnik1982, Challa1986} and the quantum spin-$1/2$ Heisenberg chain~\cite{nahum_serna_prl_2015}, wherein the ground states are emergent critical phases.

In this letter, we study a model exhibiting a variety of emergent phases and phase transitions driven by competing interactions. We focus on a 2D frustrated $q$-state clock model~\cite{Wu1982} on a square lattice with nearest-neighbor ferromagnetic (FM) coupling $J_1$ and next-nearest-neighbor antiferromagnetic (AFM) coupling $J_2$:
\begin{equation}
H = -J_1 \sum_{\braket{\mathbf{i}, \mathbf{j}}} \cos \left( \theta_{\mathbf{i}} - \theta_{\mathbf{j}} \right)
+ J_2 \sum_{\langle \braket{\mathbf{i}, \mathbf{j}} \rangle} \cos \left( \theta_{\mathbf{i}} - \theta_{\mathbf{j}} \right).
\label{eq:Hamil}
\end{equation}
Here $J_{1(2)} \geq 0$, and $\theta_{\mathbf{i}} = 2\pi k_{\mathbf{i}} / q$ with $k_{\mathbf{i}} \in {0, \dots, q-1}$ denotes the clock angle at site $\mathbf{i} \equiv (i_x, i_y)$. The Hamiltonian~\eqref{eq:Hamil} is symmetric under $\mathbb{Z}_q$ rotations $\mathcal{R}: \theta \mapsto \theta + \frac{2\pi}{q}$ and $\mathbb{Z}_2$ reflections $\mathcal{T}: \theta \mapsto -\theta$, which obey $\mathcal{T} \mathcal{R} \mathcal{T} = \mathcal{R}^{-1}$ and generate the dihedral group $D_q \cong \mathbb{Z}_q \rtimes \mathbb{Z}_2$~\cite{prakash2024classicaloriginslandauincompatibletransitions}. In addition, there are the symmetries of the square lattice. We focus on even $q$, since odd values introduce additional, nontrivial frustration.

The standard clock model ($J_2 = 0$) exhibits an emergent $U(1)$ symmetry for $q > 4$ at intermediate temperatures~\cite{Jose1977, Elitzur1979, Cardy1980, Tobochnik1982, Challa1986, Yamagata1991, Tomita2002, Lapilli2006, Matsuo_jpamg_2006, Hwang2009, Baek2010, baek_minnhagen_pre_2010, Brito2010, Borisenko2011, Borisenko2012, Ortiz2012, Baek2013, Kumano2013, Chatelain2014, Chen2017, Surungan2019, sun_prb_2019, Hong2020, Krmr2020, Li2020, miyajima_prb_2021, chen_pre_2022, tuan_pre_2022, li_prr_2022, kharinar_vojta_pre_2025,vishnu_khairnar_prb_2025}, giving rise to a critical XY-like quasi-long-range ordered (QLRO) phase, intervening between the low-temperature $\mathbb{Z}_q$-broken FM phase (denoted as $\mathbb{Z}_q^+$ FM~\footnote{Here, $\mathbb{Z}^+_q$ indicates that the spins take values that are even multiples of $\pi/q$.}) and a high-temperature paramagnetic (PM) phase. Both transitions -- PM to QLRO and QLRO to $\mathbb{Z}_q^+$ FM -- are of the Berezinskii-Kosterlitz-Thouless (BKT)~\cite{Berezinsky1970, Kosterlitz1973, Kosterlitz1974} type (c.f.~\footnote{We note that the universal nature of these two transitions for $4 < q \le 8$ has long been a subject of debate~\cite{Lapilli2006, Hwang2009, Baek2010, baek_minnhagen_pre_2010, Baek2013, Kumano2013}. However, a series of recent studies has provided compelling evidence that both transitions are of BKT type for all $q > 4$~\cite{Kumano2013, Borisenko2011, Krmr2020, Li2020, Borisenko2012, Chatelain2014, Surungan2019, chen_pre_2022, kharinar_vojta_pre_2025, vishnu_khairnar_prb_2025}.}).

For finite $J_2$, the Hamiltonian~\eqref{eq:Hamil} reduces to the well-known $J_1$-$J_2$ Ising model when $q = 2$. In this case, at low-$T$, the system exhibits an FM phase for $J_2/J_1 < 1/2$, and a 4-fold degenerate stripe-ordered phase for $J_2/J_1 > 1/2$, where spins on $45^\circ$-rotated $\sqrt{2} \times \sqrt{2}$ sublattices are anti-aligned~\cite{lopez_prb_1993, kalz_prb_2011, kalz_prb_2012, jin_prl_2012, jin_prb_2013, bobak_pre_2015, li_pre_2021, abalmasov_pre_2023, yoshiyama_pre_2023, gangat_prb_2024, lee_prb_2024, li_ptep_2024}. In the opposite limit, $q \to \infty$, the model becomes the $J_1$-$J_2$ XY model with exact $O(2)$ symmetry~\cite{Henley1987, Henley1989, henley_jpa_1998, Fernandez1991, Simon1997, Simon1997b, Loison2000, FengFeng2024}. For $J_2/J_1 < 1/2$, the system supports a low-$T$ QLRO phase that transitions to the disordered PM phase via a BKT transition. When $J_2/J_1 > 1/2$, the ground-state energy (i.e., at $T=0$) is independent of the relative orientation between the two $\sqrt{2} \times \sqrt{2}$ sublattices \cite{Henley1989, Simon1997}.
 However, at low but finite $T$, the order-by-disorder (OBD) mechanism \cite{Henley1987, Henley1989, henley_jpa_1998} lifts this continuous degeneracy, favoring a discrete Ising-like order that coexists with XY QLRO, where the relative angle between the sublattices is constrained to $0$ or $\pi$. This phase, often called a nematic (NM) vestigial phase in the literature, is referred to here as the QLRO $\oplus$ $\mathbb{Z}_2$ NM phase.
 Additionally, classical Monte Carlo (CMC) and tensor network (TN) studies have revealed another ordered phase between this QLRO $\oplus$ $\mathbb{Z}_2$ NM phase and the PM phase~\cite{Loison2000, FengFeng2024}. This intermediate phase lacks QLRO but retains $\mathbb{Z}_2$-broken order tied to the relative orientation between the sublattices, and is referred to here as the $\mathbb{Z}_2$ NM phase.

Whilst much is known about the limiting cases, $q = 2$ and $q \to \infty$, the intermediate regime remains far less explored.
Thus, in this letter, we explore the phases of the frustrated clock model for finite and even $q > 4$~\footnote{For $q=4$, the $J_1$-$J_2$ clock model is exactly equivalent to two \emph{decoupled} $J_1$-$J_2$ Ising models. This correspondence arises because the clock angles $\theta_i \in \{0, \pi/2, \pi, 3\pi/2\}$ can be mapped onto two Ising variables $\sigma_i, \tau_i \in \{\pm 1\}$ as $0 \rightarrow (+1,+1)$, $\pi/2 \rightarrow (+1,-1)$, $\pi \rightarrow (-1,-1)$, and $3\pi/2 \rightarrow (-1,+1)$. Consequently, the spin-spin interaction term transforms as $\cos(\theta_i - \theta_j) \rightarrow \tfrac{1}{2}(\sigma_i \sigma_j + \tau_i \tau_j)$.}, 
concentrating on the $J_2/J_1 > 1/2$ regime.
Strikingly, we identify an unconventional form of emergent behavior: rather than irrelevant perturbations leading to emergent continuous symmetry, frustration in combination with an RG-relevant perturbation stabilizes a distinct emergent \emph{discrete} $\mathbb{Z}_q$ order, different from the microscopic $\mathbb{Z}_q$ degrees of freedom contained in the Hamiltonian.

\begin{figure*}[htb]
    \centering
    \begin{overpic}[width=\linewidth]{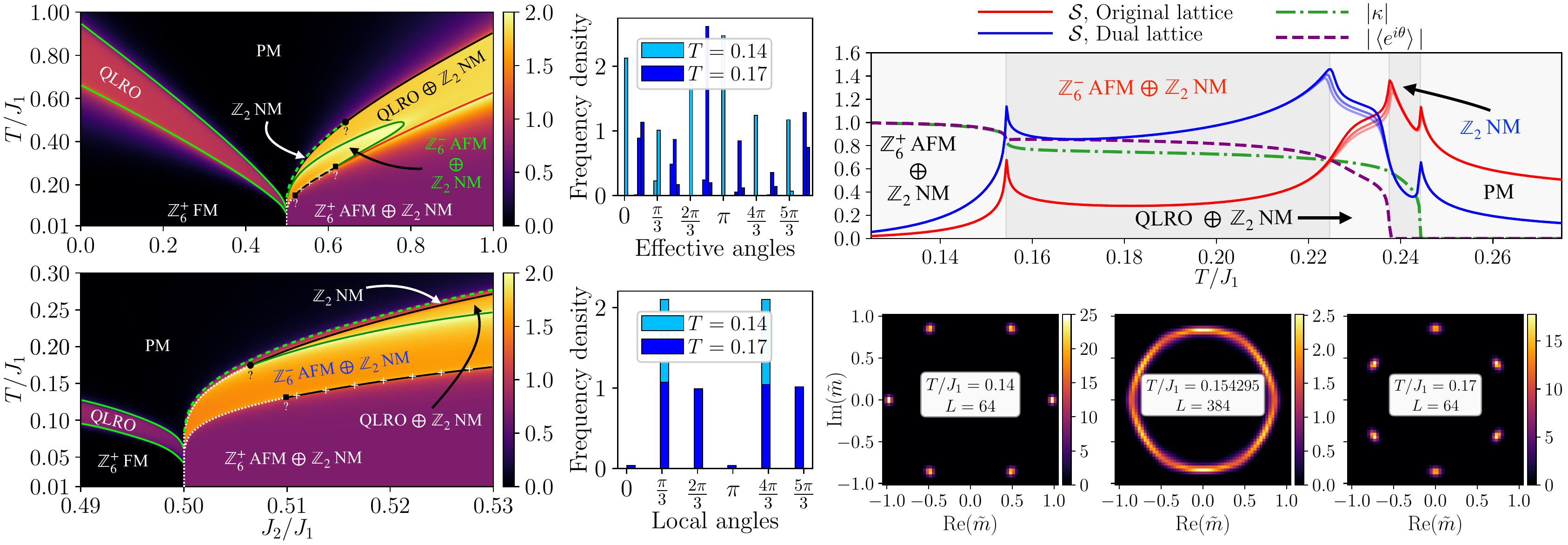}
    \put(5.5, 21.2){\textcolor{white}{(a)}} 
    \put(5.5, 15.5){\textcolor{white}{(b)}}
    \put(49.4, 31.8){(c)} 
    \put(49.4, 14.5){(d)} 
    \put(56, 32.3){(e)}
    \put(56.5, 13){\textcolor{white}{(f)}}
    \put(71.3, 13){\textcolor{white}{(g)}}
    \put(86.1, 13){\textcolor{white}{(h)}}   
    \end{overpic}
    \caption{\textbf{The phases of the system.}    
    (a)-(b) The phase diagram for $q=6$ through the lens of EE of the boundary MPS. 
    For $J_2/J_1 < 1/2$, the system follows the standard 6-state clock model: low-$T$ FM ($\mathbb{Z}_6^+$ FM), intermediate QLRO, and high-$T$ PM phases. 
    For $J_2/J_1 > 1/2$, five distinct phases emerge, including $\mathbb{Z}_6^+$ AFM $\oplus$ $\mathbb{Z}_2$ NM at low $T$ and critical QLRO $\oplus$ $\mathbb{Z}_2$ NM at intermediate $T$. 
    Furthermore, a narrow non-critical $\mathbb{Z}_2$ NM phase appears at smaller $J_2/J_1 > 1/2$ (see (b)). Most notably, a third non-critical phase ($\mathbb{Z}_6^-$ AFM $\oplus$ $\mathbb{Z}_2$ NM) emerges (the lobe in (a)), where effective $\mathbb{Z}_6$ angles shift from $2k\pi/6$ to $(2k+1)\pi/6$.
    Solid, dashed, and dotted lines indicate BKT, Ising, and first-order transitions, respectively. 
    The solid black lines with red `$+$' symbols indicate the Landau-incompatible continuous transition, see text, with
    the black squares indicating its extent.
    The black circle in (a) marks the point beyond which the Ising transition and the $\mathbb{Z}_2$ NM phase disappear within our numerical accuracy, while the black circle in (b) denotes the location where the first-order transition splits into two BKT transitions.
    The EEs are computed via isotropic CTMRG on the dual lattice with $\chi = 128$.
    (c) Histograms of the effective clock angles
    obtained with CMC for a lattice of linear size $L=64$, at $\mathbb{Z}_6^+$ AFM $\oplus$ $\mathbb{Z}_2$ NM ($T=0.14$) and $\mathbb{Z}_6^-$ AFM $\oplus$ $\mathbb{Z}_2$ NM ($T=0.17$) phases with $J_2/J_1 = 0.52$.
    (d) Histograms of the local clock angles 
    from a single CMC snapshot, corresponding to the same parameter points as in (c).
    (e) Observables for $J_2/J_1 = 0.52$ as functions of $T/J_1$, including EEs (in both original and dual lattices), and the order parameters $\braket{e^{i\theta}}$ and $\kappa$. Here, anisotropic CTMRG is used for both the models with $\chi = 128$. The EEs are also computed with $\chi = 96$ and $80$ that shows strong $\chi$ dependence in the critical QLRO $\oplus$ $\mathbb{Z}_2$ NM phase.
    (f)-(h) 
    Frequency distribution of the complex order parameter $\tilde{m}$ in the complex plane, obtained from $2.5 \times 10^6$ CMC snapshots for system sizes $L = 64$ and $384$. Results are shown for three representative points: (f) the $\mathbb{Z}_q^+$ AFM $\oplus$ $\mathbb{Z}_2$ NM phase ($T/J_1 = 0.14$), (g) the Landau-incompatible continuous transition ($T/J_1 = 0.154295$), and (h) the $\mathbb{Z}_q^-$ AFM $\oplus$ $\mathbb{Z}_2$ NM phase ($T/J_1 = 0.17$), with $J_2/J_1 = 0.52$ fixed. At the transition, the distribution becomes rotationally invariant, reflecting the emergent $O(2)$ symmetry. The apparent hexagonal pattern in (g) is a finite-size artifact that diminishes with increasing system size.
    }
    \label{fig:phase}
\end{figure*}

 \paragraph{The numerical methods.--} We use two complementary numerical methods. First, we perform standard CMC simulations~\cite{book_newman, metropolis_rosenbluth_jcp_53}. Due to frustration, efficient cluster updates are unavailable, so we rely on single-spin-flip Metropolis updates, which suffer from long autocorrelation times -- especially in critical regimes. This limits the resolution of phase boundaries and narrow phases. Nonetheless, deep inside each phase, CMC remains effective for extracting key properties~\cite{sm}.

 To complement the CMC analysis, we employ TN methods~~\cite{Nishino1996, Nishino1997, Corboz2014, Fishman2018, Levin2007, Gu2008, Evenbly2015, Ran2020, Okunishi2022, Xiang2023}, specifically variants of the corner transfer matrix renormalization group (CTMRG)~\cite{Nishino1996, Nishino1997, Corboz2014, Fishman2018}. The partition function for Eq.~\eqref{eq:Hamil} can be written as a 2D TN, which we contract efficiently in the thermodynamic limit using anisotropic CTMRG~\cite{Corboz2014, Fishman2018}. To capture $\mathbb{Z}_2$ nematic order, we use a $4 \times 4$ unit cell (see~\cite{sm}).

Alternatively, we analyze an equivalent model on the dual lattice by placing spins on bonds: 
$\sigma_{(\mathbf{i}, \hat{x})} := (-1)^{i_x+i_y}(\theta_{\mathbf{i} + \hat{x}} - \theta_{\mathbf{i}})$ and $\sigma_{(\mathbf{i}, \hat{y})} := (-1)^{i_x+i_y}(\theta_{\mathbf{i}} - \theta_{\mathbf{i}+\hat{y}})$~\cite{Chen2017, Li2020}.
The dual Hamiltonian becomes
\begin{align}
H_D = -J_1 \sum_{\mathbf{b}} \cos \sigma_{\mathbf{b}} + \frac{J_2}{2} \sum_{\langle \mathbf{b}, \mathbf{b}' \rangle} \cos(\sigma_{\mathbf{b}} + \sigma_{\mathbf{b}'}),
\label{eq:Hamil_dual}
\end{align}
with the local constraint $\cos(\sigma_{(\mathbf{i}, \hat{x})} + \sigma_{(\mathbf{i}, \hat{y})} + \sigma_{(\mathbf{i} + \hat{x}, \hat{y})} + \sigma_{(\mathbf{i} + \hat{y}, \hat{x})}) = 1$ at each site-index $\mathbf{i}$. The bond variables $\sigma_{\mathbf{b}}$ take values that are even multiples of $\pi/q$, as before. This dual representation simplifies the partition function~\cite{sm}, allowing efficient contraction via the original isotropic CTMRG algorithm~\cite{Nishino1996, Nishino1997} using a single-site unit cell.

\paragraph{The phase diagram.--} Figure~\ref{fig:phase}(a) and (b) show the phase diagram for $q=6$. In the end matter (EM), we also provide the results for $q=8$, 10, and 12, 
indicating that the qualitative structure remains similar for any finite, even $q > 4$.
These phase diagrams are based on the entanglement entropy (EE) $\mathcal{S}$ of the boundary matrix-product state (MPS) (see~\cite{sm}), computed via isotropic CTMRG for the dual model~\eqref{eq:Hamil_dual} with MPS bond dimension $\chi = 128$.

For $J_2/J_1 < 1/2$, the phase diagram closely mirrors that of the standard $q$-state clock model, featuring three familiar phases ($\mathbb{Z}_q^+$ FM, QLRO, and PM). In contrast, for $J_2/J_1 > 1/2$, the diagram becomes significantly richer, exhibiting five distinct phases and various transitions between them. As in the XY limit, the ground-state energy (i.e., at $T=0$) is degenerate with respect to the relative angle between the two $\sqrt{2} \times \sqrt{2}$ sublattices. Within each sublattice, spins prefer an AFM arrangement that can now establish true long-range order due to the discreteness of the $\mathbb{Z}_q$ symmetry. At low but finite $T$, the OBD kicks in and restricts the relative angle either to $0$ or $\pi$, resulting in a stripe-ordered phase with $2q$-fold degeneracy. We refer to this as the $\mathbb{Z}_q^+$ AFM $\oplus$ $\mathbb{Z}_2$ NM phase.

At intermediate $T$ and for $J_2/J_1 > 1/2$, an emergent $U(1)$ symmetry -- reminiscent of the $J_2 = 0$ case -- renders the effective spins rotationally invariant. As a result, true AFM order within the sublattices is suppressed by the Mermin–Wagner theorem~\cite{Mermin1966, Coleman1973}, yielding a critical QLRO state. However, the spatial $\mathbb{Z}_2$ symmetry governing the relative sublattice orientation remains broken, resulting in the QLRO $\oplus$ $\mathbb{Z}_2$ NM phase, akin to the low-$T$ regime of the $J_1$-$J_2$ XY model. Additionally, we identify a narrow, non-critical $\mathbb{Z}_2$ NM phase in the small $J_2/J_1 > 1/2$ regime, situated between the QLRO $\oplus$ $\mathbb{Z}_2$ NM and PM phases (see Fig.~\ref{fig:phase}(b)).

\paragraph{Emergent $\mathbb{Z}_q$ spins.--} 
Most intriguingly, at intermediate $T$ and lower values of $J_2/J_1 > 1/2$, we uncover another non-critical phase with $2q$-fold degeneracy and stripe order (see the lobe in Fig.~\ref{fig:phase}(a)). In this regime, the effective spin degrees of freedom -- relevant in the low-energy infrared limit -- depart from those in the microscopic model (Eq.~\eqref{eq:Hamil}), instead taking values that are \textit{odd multiples of $\pi/q$}. This emergence of forbidden degrees of freedom in the infrared limit, induced by competing interactions, is a central result of this work. We refer to this phase as the $\mathbb{Z}_q^-$ AFM $\oplus$ $\mathbb{Z}_2$ NM~\footnote{Here, $\mathbb{Z}_q^-$ signifies that the effective spins in the infrared limit take odd multiples of $\pi/q$.} phase.

We confirm this emergent behavior by computing the staggered magnetization on one of the $\sqrt{2} \times \sqrt{2}$ sublattices for each CMC snapshot:
\begin{equation}
\tilde{m} = \frac{1}{L^2/2} \sum_{i_x, i_y} (-1)^{i_x+i_y} \exp(i \theta_{(i_x, i_y)}),
\label{eq:afm_mag}
\end{equation}
where $i_{x(y)}$ runs over the rotated sublattice and $L$ is the linear size of the system. Figure~\ref{fig:phase}(c) shows histograms of the effective clock angle, i.e., the phase of the complex number $\tilde{m}$ from CMC snapshots with varied initializations, comparing the $\mathbb{Z}_q^{\pm}$ AFM $\oplus$ $\mathbb{Z}_2$ NM phases. In the $\mathbb{Z}_q^+$ phase, the angles cluster around $2k\pi/q$ for $k = 0, 1, ..., q-1$, while in the $\mathbb{Z}_q^-$ phase, they center at $(2k+1)\pi/q$, indicating the emergent effective degrees of freedom (see also Fig.~\ref{fig:phase}(f) and (h)). Figure~\ref{fig:phase}(d) presents histograms of local clock angles for a single CMC snapshot for one of the $\sqrt{2} \times \sqrt{2}$ sublattices. 
In the $\mathbb{Z}_q^+$ phase, two prominent peaks separated by $\pi$ reflect standard AFM order. In contrast, the $\mathbb{Z}_q^-$ phase shows two pairs of peaks of nearly equal frequencies. Within each pair the peaks are separated by $2\pi/q$, and the pairs themselves are spaced by $\pi$ relative to their midpoint -- highlighting AFM order built from emergent $(2k+1)\pi/q$ spin states.

\paragraph{Order parameters and observables.--} 
To characterize the various phases and locate transition boundaries, we examine several observables and order parameters (see Fig.~\ref{fig:phase}(e)). First, we compute the EE for both the original and dual systems using anisotropic CTMRG with $4 \times 4$ and $2 \times 2$ unit cells, respectively~\footnote{See the supplemental material~\cite{sm} for details on the unit cell choice for the dual system.}. Kinks in the EE effectively signal phase boundaries, including BKT transitions where analyzing both the original and dual systems proves useful~\cite{Chen2017, Li2020}. $\mathbb{Z}_q$ symmetry-breaking is detected via the non-zero $\mathbb{Z}_q$ order parameter $\braket{e^{i\theta}}$ at a representative CTMRG site. The $\mathbb{Z}_2$ NM order is identified using the nematic order parameter:
$\kappa = \frac{1}{4} \langle \cos\left(\theta_{\mathbf{i}} - \theta_{\mathbf{i} + \hat{x}}\right) + \cos\left(\theta_{\mathbf{i} + \hat{y}} - \theta_{\mathbf{i} + \hat{x} + \hat{y}}\right)
- \cos\left(\theta_{\mathbf{i}} - \theta_{\mathbf{i} + \hat{y}}\right)  - \cos\left(\theta_{\mathbf{i} + \hat{x}} - \theta_{\mathbf{i} + \hat{x} + \hat{y}}\right) \rangle$.
This quantity approaches $\kappa \sim \pm 1$ deep within the nematic phases, as shown in Fig.~\ref{fig:phase}(e).
In Tab.~\ref{tab:orders}, we summarize all the phases present in the system classified according to their symmetry properties.

\begin{table}[htb]
    \centering
    \begin{tabular}{|c|c|c|c|}
    \hline
     \textbf{Phases}   &  \textbf{$\mathbb{Z}_q$ symm.} &  \textbf{is critical} &  \textbf{$\mathbb{Z}_2$ symm.} \\
      & & ? & \textbf{(spatial)} \\
      \hline
      \hline
      $\mathbb{Z}_q^+$ FM & Broken & No & Unbroken \\
      QLRO & -- & Yes & Unbroken \\
      $\mathbb{Z}_q^+$ AFM $\oplus$ $\mathbb{Z}_2$ NM & Broken & No & Broken \\
      QLRO $\oplus$ $\mathbb{Z}_2$ NM & -- & Yes & Broken \\
      $\mathbb{Z}_2$ NM & -- & No & Broken \\
      $\mathbb{Z}_q^-$ AFM $\oplus$ $\mathbb{Z}_2$ NM & Broken (!) & No & Broken \\
      PM & Unbroken & No & Unbroken \\
      \hline
    \end{tabular}
    \caption{\textbf{Phases of the system.} Summary of all phases and their corresponding symmetry attributes. The exclamation mark for the $\mathbb{Z}_q^-$ AFM $\oplus$ $\mathbb{Z}_2$ NM highlights that the effective $\mathbb{Z}_q$ spins are emergent.
    The critical nature of the respective QLRO phases, is confirmed by critical EE scaling~\cite{callan_geometric_1994, vidal_PRL_2003, calabrese_entanglement_2004}, see~\cite{sm}.
    }
    \label{tab:orders}
\end{table}

\begin{figure}[htb]
    \centering
    \begin{overpic}[width=\linewidth]{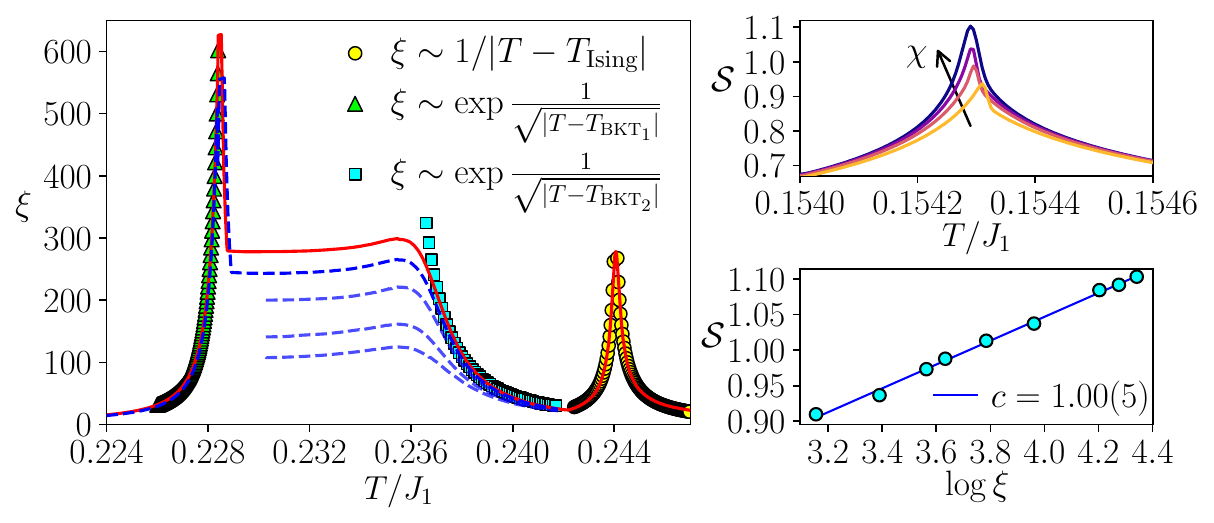}
    \put(9.5, 38){(a)} 
    \put(67, 38){(b)}
    \put(67, 17.5){(c)}
    \end{overpic}
    \caption{\textbf{Characterizing critical transitions.}
    (a) Correlation length $\xi$ in the dual system for MPS bond dimension $\chi=640$ (red solid line) at $J_2/J_1=0.52$, as $T/J_1$ is varied across $\mathbb{Z}_q^-$ AFM $\oplus$ $\mathbb{Z}_2$ NM, QLRO $\oplus$ $\mathbb{Z}_2$ NM,  $\mathbb{Z}_2$ NM, and PM phases. Dashed blue lines show $\xi$ for $\chi = 192, 256, 384, 512$. Expected critical scaling relations for BKT and Ising transitions are verified by fits near the transitions. 
    (b) EE across the Landau-incompatible continuous transition between the $\mathbb{Z}_q^{\pm}$ AFM $\oplus$ $\mathbb{Z}_2$ NM phases at $J_2/J_1=0.52$ from anisotropic CTMRG with $\chi = 48, 64, 80, 112$. 
    (c) Finite-entanglement scaling of the EE peak for different bond dimensions $\chi \in [40:112]$ at the Landau-incompatible transition yields central charge $c=1$.}
    \label{fig:characterize}
\end{figure}

\paragraph{Phase transitions.--}
The transitions involving both the critical phases are of the BKT type, 
where the correlation length $\xi$ diverges exponentially as $\xi \sim \exp\left(1/\sqrt{|T-T_{\text{BKT}}|}\right)$, with $T_{\text{BKT}}$ being the transition temperature. We confirm this by fitting the $\chi$-dependent correlation length (see Fig.~\ref{fig:characterize}(a)), computed directly from the fixed-point MPS tensors~\cite{sm}. In contrast, the $\mathbb{Z}_2$ NM $\leftrightarrow$ PM transition belongs to the Ising universality class, with $\xi$ scaling as $\xi \sim |T-T_{\text{Ising}}|^{-\nu}$ with the universal exponent $\nu$ being $1$ and $T_{\text{Ising}}$ being the transition temperature (Fig.~\ref{fig:characterize}(a)).

The $\mathbb{Z}_q^\pm$ AFM $\oplus$ $\mathbb{Z}_2$ NM phases break mutually incompatible symmetries. 
Within the Landau paradigm~\cite{SenthilDQCdoi:10.1126/science.1091806,SenthilDQCreview_doi:10.1142/9789811282386_0014}, such phases can be separated either by a first-order transition (dotted lines in Fig.~\ref{fig:phase}(a) and (b); see \cite{sm} for details) or by an intervening phase (here the intervening QLRO $\oplus$ $\mathbb{Z}_2$ NM phase~\footnote{The QLRO $\oplus$ $\mathbb{Z}_2$ NM phase itself, on the other hand, is not within the Landau paradigm.}, Fig.~\ref{fig:phase}(a)). 
Remarkably, we also uncover a line of \textit{Landau-incompatible} continuous transitions~\cite{prakash2024classicaloriginslandauincompatibletransitions} between them (solid lines with `$+$' symbols in Fig.~\ref{fig:phase}(a) and (b)).

Across this transition, EE exhibits characteristic bond-dimension-dependent peaks (Fig.~\ref{fig:characterize}(b)). To characterize this criticality, we perform finite-entanglement scaling~\cite{callan_geometric_1994, vidal_PRL_2003, calabrese_entanglement_2004}: $\mathcal{S} = (c/6) \log \xi + b$, where $c$ is the central charge of the underlying conformal field theory (CFT) and $b$ is a non-universal constant, applied to EE peak values for different bond dimensions~\cite{Kjall2013, Tsitsishvili2022}. The resulting scaling (Fig.~\ref{fig:characterize}(c)) indicates that the criticality is governed by a CFT with $c=1$.
Moreover, as shown in Fig.~\ref{fig:phase}(g), the system at this transition exhibits an emergent $O(2)$ symmetry as the effective clock spins become rotationally invariant.
We note that determining the extent of this critical line, indicated by the black squares in Fig.~\ref{fig:phase}(a) and (b), is numerically challenging; these squares therefore represent approximate end points.

The direct transition between $\mathbb{Z}_q^-$ AFM  $\oplus$ $\mathbb{Z}_2$ NM and $\mathbb{Z}_2$ NM phases is first-order (see~\cite{sm}), except for the end-point of the first-order transition line (marked by black circle in Fig.~\ref{fig:phase}(b)), where the transition is again continuous. We note here that numerical characterization of this continuous transition, at the end of first-order transition line, is difficult within our numerical accuracy as nearby states have large correlation lengths.

\begin{figure}
    \centering
    \includegraphics[width=0.8\linewidth]{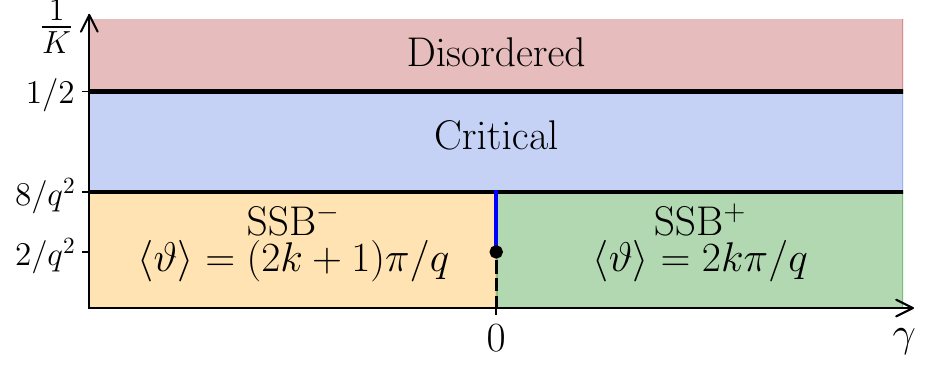}
    \caption{\textbf{Phase diagram of the effective field theory.} 
    The theory in Eq.~\eqref{eq:S} with $\gamma' > 0$ predicts a phase diagram comprising a disordered phase, a critical phase, and two Landau-incompatible spontaneously symmetry-broken (SSB$^\pm$) phases, each possessing $q$ distinct vacua. Horizontal solid lines indicate BKT transitions, while vertical solid and dashed lines correspond to second-order transitions (with continuously varying critical exponents) and first-order transitions, respectively.}
    \label{fig:field theory phase diagram}
\end{figure}
\paragraph{Effective field theory.--} 
Various aspects of the phase diagram can be understood using an effective field theoretic analysis by mapping onto a parent Gaussian field theory~\cite{Jose1977,prakash2024classicaloriginslandauincompatibletransitions} in presence of instabilities, described by the Euclidean action:
\begin{multline}
    S \approx  \frac{K}{2 \pi} \int d^2x  \left( \nabla \vartheta\right)^2   - h \int d^2x   \cos (\phi)  \\ -\gamma \int d^2x    \cos \left(q \vartheta\right) -\gamma' \int d^2x   \cos \left(2q \vartheta\right) + \ldots   , \label{eq:S}
\end{multline}
where $\vartheta$ and $\phi$ are dual variables. We assume that all coupling constants, $h, \gamma, \gamma', \ldots$, are small, and $\gamma'>0$. The phase diagram of \cref{eq:S}, schematically shown in \cref{fig:field theory phase diagram}, can be determined by identifying the most relevant operator (in the RG sense) that would dominate long-distance physics~\footnote{Recall that an operator is relevant if its scaling dimension is smaller than the Euclidean dimensions (2 in our case). If multiple operators are relevant, the dominant operator is the one with smaller scaling dimensions.}. This, in turn, is determined by tracking the scaling dimensions of the various primary operators, which can be expressed in terms of the Luttinger parameter $K$ as $[\cos \phi]=K$, $[\cos(q \vartheta)]=q^2/(4K)$ and $[\cos (2q \vartheta)]=q^2/K$. For $q>4$, a critical phase is obtained for $2 \le K \le q^2/8$ where no operator is relevant. This phase undergoes two BKT transitions to (a) the disordered phase at $K=2$ when $\cos \phi$ becomes relevant and (b) symmetry-broken phases at $K=q^2/8$ with $q$ distinct vacua, as shown in \cref{fig:field theory phase diagram}. The symmetry-broken phases are distinct for even $q$~\cite{prakash2024classicaloriginslandauincompatibletransitions} and are separated at $\gamma=0$ by a continuous transition for $q^2/8<K<q^2/2$ with varying critical exponents and by a first-order transition for $K>q^2/2$ when $\cos(2 q \vartheta)$ becomes relevant.

The effective field theory \eqref{eq:S} reproduces different limits of the full phase diagram shown in \cref{fig:phase}. The correspondence between the phases of the field theory and the microscopic model is shown in \cref{tab:phase map} in two regimes. For $J_2/J_1<1/2$, we can interpret the field $\vartheta$ to be a coarse-grained version of the microscopic $q$-fold rotors $\theta_\mathbf{j}$ and carrying all their symmetry charges, but elevated to full angular freedom $\theta \in[0,2\pi]$. The terms $\gamma \cos(q\vartheta), \gamma' \cos(2q \vartheta), \ldots$ can be interpreted as a soft implementation of the $q$-fold anisotropy, and thus $\gamma, ~\gamma'$ are restricted to positive values to generate the correct set of $q$ angular values. The field theory can be derived by replacing the original clock model with an anisotropic XY model~\cite{Jose1977,prakash2024classicaloriginslandauincompatibletransitions} and following the recipe of Villain~\cite{Villain1975theory}. 

For $J_2/J_1>1/2$, our main regime of interest, the field $\vartheta$ corresponds to fluctuations \emph{atop} the $\mathbb{Z}_2$ NM order arising from OBD, and are not directly related to the lattice rotors. The coupling $\gamma$ therefore is allowed to take positive and negative values. The former generates the exotic $\mathbb{Z}_q^-$ AFM order with seemingly prohibited expectation values. Unlike the usual case where an irrelevant operator yields emergent symmetry~\cite{Witten2018}, here the discrete $\mathbb{Z}_q^-$ degrees of freedom emerge because $\cos(q\vartheta)$ becomes relevant.
The PM phase for $J_2/J_1>1/2$ is obtained when the $\mathbb{Z}_2$ NM order is destroyed independently and is not captured by the field theory content of \cref{eq:S}.

\begin{table}[]
\begin{tabular}{|c|c|c|c|c|}
\hline
Regimes
 &
  Disordered &
  Critical &
  SSB$^+$ &
  SSB$^-$ \\ 
  \hline
  \hline
$J_2/J_1 < 1/2$ &
  PM &
  QLRO &
  $\mathbb{Z}_q^+$ FM &
  - \\
  \hline
$J_1/J_2>1/2$ &
  $\mathbb{Z}_2$ NM &
  \begin{tabular}[c]{@{}c@{}}QLRO \\ $\oplus$ \\ $\mathbb{Z}_2$ NM\end{tabular} &
  \begin{tabular}[c]{@{}c@{}}$\mathbb{Z}_q^+$ AFM \\ $\oplus$ \\ $\mathbb{Z}_2$ NM\end{tabular} &
  \begin{tabular}[c]{@{}c@{}}$\mathbb{Z}_q^-$ AFM \\ $\oplus$ \\ $\mathbb{Z}_2$ NM\end{tabular} \\ \hline
\end{tabular}
\caption{Correspondence between phases of the effective field theory and those of the microscopic model in various limiting cases.} 
\label{tab:phase map}
\end{table}

Although we do not have a microscopic derivation of the field theory in the $J_2/J_1>1/2$ limit, the postulated effective theory is insightful as it reproduces universal aspects of the phase diagram including the line of continuous transition between the $\mathbb{Z}_q^\pm$ AFM orders.  As explained in Ref.~\cite{prakash2024classicaloriginslandauincompatibletransitions}, these two phases are \emph{Landau-incompatible} and a direct continuous transition is the classical equivalent of `deconfined quantum criticality'~\cite{SenthilDQCdoi:10.1126/science.1091806,SenthilDQCreview_doi:10.1142/9789811282386_0014} (see also the EM). Our field theory predicts that at this transition the system, in the long-wavelength limit, is described by the critical Gaussian field theory with central charge $c=1$ (see Fig.~\ref{fig:characterize}(c)) exhibiting  an emergent $O(2)$ symmetry (see Fig.~\ref{fig:phase}(g)), as confirmed by our numerics.

On the other hand, the first-order transition and their critical end-point (black circle in Fig.~\ref{fig:phase}(b)) between the $\mathbb{Z}_q^-$ AFM $\oplus$ $\mathbb{Z}_2$ NM and $\mathbb{Z}_2$ NM phases lie outside the scope of Eq.~\eqref{eq:S}. Our numerics also indicate that both $\mathbb{Z}_2$ NM order and QLRO vanish together with increasing $T$ once $J_2/J_1 \gtrsim 0.65$ (black circle in Fig.~\ref{fig:phase}(a)), a feature not captured by the current effective theory. A microscopic derivation of the field theory for $J_2/J_1 > 1/2$ is therefore highly desirable. In particular, an explicit coupling between Eq.~\eqref{eq:S} and the Ising sector governing $\mathbb{Z}_2$ NM order may provide a field-theoretic understanding of the phase transitions between the $\mathbb{Z}_q^-$ AFM $\oplus$ $\mathbb{Z}_2$ NM and $\mathbb{Z}_2$ NM phases, and may resolve the outstanding question of the simultaneous loss of QLRO and NM order~\cite{granato_kosterlitz_prl_1991, lee_granato_prb_1991, nightingale_granato_prb_1995}.

\paragraph{Conclusions.--} In this letter, we have unraveled the phases and phase transitions of the frustrated $J_1$-$J_2$ clock model by laying recourse to corner transfer matrix renormalization group techniques supplanted by classical Monte-Carlo simulations. We remarkably show the existence of a discretely ordered emergent phase (for even $q>4$) that contravenes the existing wisdom behind emergence. Furthermore, we show that the system, apart from hosting traditional phase transitions such as Ising, BKT, and first-order,  also features unconventional Landau-incompatible deconfined transitions. Our numerical findings are supported via an effective field-theoretical framework. 
We note that experimental realizations of such frustrated clock models and observation of their unconventional emergent behavior are within reach using platforms like Josephson‑junction chains~\cite{Wauters_2025}.

\acknowledgements
We thank Z. Bacciconi and M. Dalmonte for useful discussions.
P.K.V. acknowledges the use of Aqua cluster at IIT Madras. P.K.V. and R.N. acknowledge funding from the Center for Quantum Information Theory in Matter and Spacetime, IIT Madras and from the Department of Science and Technology, Govt. of India, under Grant No. DST/ICPS/QuST/Theme-3/2019/Q69, as well as support from the Mphasis F1 Foundation via the Centre for Quantum Information, Communication, and Computing (CQuICC).
A.P. and T.C. are grateful to the International Centre for Theoretical Sciences (ICTS) for participating in the discussion meeting - (QM100) A Hundred Years of Quantum Mechanics (code: ICTS/qm100/2025/01).
T.C. acknowledges the support by the Young Faculty Initiation Grant (NFIG) at IIT Madras (Project No. RF24250775PHNFIG009162) and 
from the Anusandhan National Research Foundation (ANRF), India via the Prime Minister Early Career Research Grant ANRF/ECRG/2024/001198/PMS.
T.C. thanks the Polish high-performance computing infrastructure PLGrid (HPC Center: ACK Cyfronet AGH) for providing computer facilities within computational grant no.  PLG/2024/017289. 
We further acknowledge National Supercomputing Mission (NSM) for providing computing
resources of `PARAM RUDRA' at the P. G. Senapathy Center For Computer Resources, Play Field Ave, Indian Institute of Technology Madras, Tamil Nadu 600036, which is implemented by C-DAC and supported by the Ministry of Electronics and Information Technology (MeitY) and Department of Science and Technology (DST), Government of India.
The CTMRG calculations have been performed using ITensors.jl library~\cite{ITensor2022, ITensor2022codebase}.

\bibliography{main.bbl}

\onecolumngrid
\section*{End Matter}
\twocolumngrid

\subsection*{The $J_1$-$J_2$ clock model for $q > 6$}

\begin{figure*}[htb]
    \centering
    \begin{overpic}[width=1\linewidth]{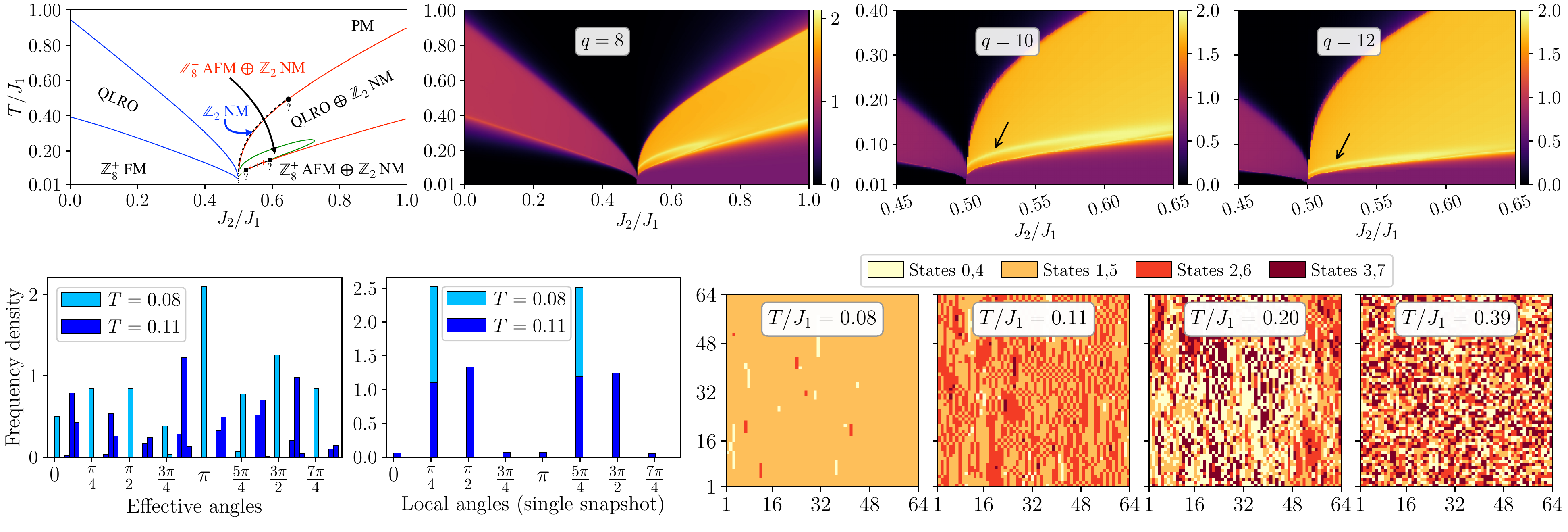}
    \put(5, 31.5){(a)}
    \put(30, 23){\textcolor{white}{(b)}} 
    \put(19, 14){(c)}
    \put(40.5, 14){(d)}
    \put(50, 16){(e)}
    \put(57.6, 31.5){\textcolor{white}{(f)}} 
    \put(79.3, 31.5){\textcolor{white}{(g)}} 
    \end{overpic}
    \caption{\textbf{Phases of the system for $q > 6$.}
    (a) The schematic phase diagram of the system for $q=8$ in the $(J_2/J_1, T/J_1)$-plane. All other details remain the same as in Fig.~\ref{fig:phase}(a). The schematic diagram is constructed based on the EE pattern, panel (b),  obtained from isotropic CTMRG for the dual system with MPS bond dimension $\chi=128$.
    (c) Histograms of the effective clock angles for $q = 8$ computed via CMC, at the $\mathbb{Z}_8^+$ AFM $\oplus$ $\mathbb{Z}_2$ NM ($T=0.08$) and the $\mathbb{Z}_8^-$ AFM $\oplus$ $\mathbb{Z}_2$ NM ($T=0.11$) phases $(J_2/J_1 = 0.52)$, for a system of size $L=64$.
    (d) Histograms of the local clock angles from a single CMC snapshot, corresponding to the same parameter points as in (c).
    (e) Real-space snapshots of clock states $k \in \{0,1,\dots,q-1\}$ for $q=8$ obtained from CMC for a system of linear size $L=64$.
    Snapshots are shown for the phases: 
    $\mathbb{Z}_q^+$ AFM $\oplus$ $\mathbb{Z}_2$ NM ($T/J_1 = 0.08$), 
    $\mathbb{Z}_q^-$ AFM $\oplus$ $\mathbb{Z}_2$ NM ($T/J_1 = 0.11$), 
    QLRO $\oplus$ $\mathbb{Z}_2$ NM ($T/J_1 = 0.20$), 
    and PM ($T/J_1 = 0.39$). $J_2/J_1$ is fixed to $0.52$.
    For clarity of visualization, the AFM order in the two $\sqrt{2} \times \sqrt{2}$ sublattices has been removed by rotating the spins on odd sublattice sites by $\pi$ (see text).
    (f) The pattern of EE for $q=10$ obtained from isotropic CTMRG for the dual system with MPS bond dimension $\chi=128$. (g) Same as in panel (f), but for $q=12$. The arrows in panels (f) and (g) indicate the phase with emergent $\mathbb{Z}_q^-$-spins.}
    \label{fig:phase_8}
\end{figure*}

In the main text, we have analyzed the properties of the various phases and phase transitions in the $J_1$-$J_2$ clock model with a fixed $q=6$. Here we show that the qualitative structure of the phase diagram, along with the quantitative characteristics of the phases and transitions, remains consistent for larger values of $q$.

To confirm this generality, we present numerical results for $q=8$, $10$, and $12$. In Fig.~\ref{fig:phase_8}(a), we show the schematic phase diagram for the system with $q=8$. This diagram is constructed based on the pattern of EE, as shown in Fig.~\ref{fig:phase_8}(b), obtained using isotropic CTMRG for the dual system~\eqref{eq:Hamil_dual}. The phase diagram closely resembles the $q=6$ one, with all phases listed in Tab.~\ref{tab:orders} observed for $q=8$. However, the phase featuring emergent $\mathbb{Z}^-_8$-spins, i.e., the $\mathbb{Z}_8^-$ AFM $\oplus$ $\mathbb{Z}_2$ NM phase, diminishes in size as $q$ is increased.

\begin{figure}
    \centering
    \includegraphics[width=0.7\linewidth]{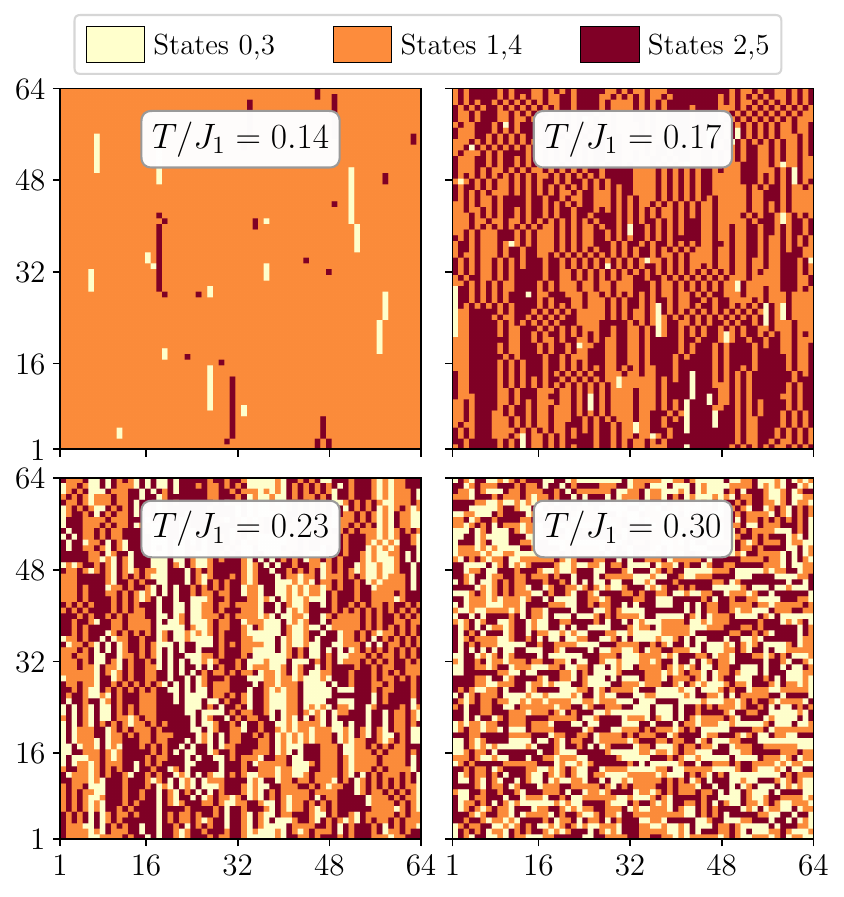}
    \caption{\textbf{Real-space snapshots of clock states for $q=6$.}
    Snapshots are shown for the phases: 
    $\mathbb{Z}_q^+$ AFM $\oplus$ $\mathbb{Z}_2$ NM ($T/J_1 = 0.14$), 
    $\mathbb{Z}_q^-$ AFM $\oplus$ $\mathbb{Z}_2$ NM ($T/J_1 = 0.17$), 
    QLRO $\oplus$ $\mathbb{Z}_2$ NM ($T/J_1 = 0.23$), 
    and PM ($T/J_1 = 0.3$). $J_2/J_1$ is fixed to $0.52$. All other details are same as in Fig.~\ref{fig:phase_8}(e).
    }
    \label{fig:snap_6}
\end{figure}

To illustrate the behavior of the emerging spin degrees of freedom, we compute the staggered magnetization $\tilde{m}$ (Eq.~\eqref{eq:afm_mag}) within a single $\sqrt{2} \times \sqrt{2}$ sublattice. In Fig.~\ref{fig:phase_8}(c), we present the histogram of effective clock angles, i.e., the phases of $\tilde{m}$, obtained from multiple CMC snapshots. As before, in the $\mathbb{Z}_8^+$ AFM $\oplus$ $\mathbb{Z}_2$ NM phase ($T=0.08$), these angles take values corresponding to even multiples of $\pi/8$. In contrast, within the emerging $\mathbb{Z}_8^-$ AFM $\oplus$ $\mathbb{Z}_2$ NM phase ($T=0.11$), the angles are centered around odd multiples of $\pi/8$ -- values that are forbidden by the microscopic Hamiltonian~\eqref{eq:Hamil}.

Similar to the $q=6$ case, this emergence can be understood by examining the distribution of local clock angles on a $\sqrt{2} \times \sqrt{2}$ sublattice from a single CMC snapshot, as shown in Fig.~\ref{fig:phase_8}(d). In the $\mathbb{Z}_8^+$ AFM $\oplus$ $\mathbb{Z}_2$ NM phase, the histogram exhibits two peaks separated by $\pi$, confirming the AFM nature within the $\sqrt{2} \times \sqrt{2}$ sublattice. In contrast, for the $\mathbb{Z}_8^-$ AFM $\oplus$ $\mathbb{Z}_2$ NM phase, the histogram reveals two pairs of (major) peaks of nearly equal frequencies. Within each pair, the peaks are separated by $\pi/4$, while the pairs themselves are spaced by $\pi$ relative to their midpoint. This  explains the emergence of effective angles corresponding to odd multiples of $\pi/8$, which maintain an AFM order within the $\sqrt{2} \times \sqrt{2}$ sublattice.

To further illustrate these behaviors, Fig.~\ref{fig:phase_8}(e) shows real-space snapshots of the clock states in the four phases: $\mathbb{Z}_q^+$ AFM $\oplus$ $\mathbb{Z}_2$ NM, $\mathbb{Z}_q^-$ AFM $\oplus$ $\mathbb{Z}_2$ NM, QLRO $\oplus$ $\mathbb{Z}_2$ NM, and PM for $q=8$.  For clearer visualization, we remove the AFM component in the two $\sqrt{2}\times \sqrt{2}$ sublattices by rotating the spins on odd sublattice sites by $\pi$ (i.e., mapping them to the opposite clock states):
\begin{align}
k & \rightarrow k, && \text{for even } (i_x + i_y), \nonumber \\
k & \rightarrow (k + q/2) \bmod q, && \text{for odd } (i_x + i_y),
\end{align}
where $k \in {0,1,\dots,q-1}$ labels the clock states, and $(i_x,i_y)$ are the $\sqrt{2}\times\sqrt{2}$ sublattice coordinates.
This transformation effectively converts the AFM order into an FM one. 
Figure~\ref{fig:snap_6} shows the real-space snapshots in different phases for $q=6$.
These real-space snapshots clearly reveal the emergence of the new $\mathbb{Z}_q$ order. In the standard $\mathbb{Z}_q^+$ AFM $\oplus$ $\mathbb{Z}_2$ NM phase, only one pair of states separated by $\pi$ is visible, whereas in the $\mathbb{Z}_q^-$ AFM $\oplus$ $\mathbb{Z}_2$ NM phase two distinct pairs of states appear, separated by $\Delta k=1$. In the  long-wavelength limit this structure corresponds to effective half-integer clock states, with spin orientations at odd multiples of $\pi/q$.

Our field-theoretical framework predicts that the phase with emergent $\mathbb{Z}_q^-$ degrees of freedom should persist for all finite even $q > 4$, with its extent expected to decrease as $q$ increases, ultimately vanishing in the XY limit ($q \to \infty$). The entanglement entropy (EE) patterns for $q=10$ and $q=12$ shown in Fig.~\ref{fig:phase_8}(f) and (g) support this scenario, as the domain of the emergent phase indeed shrinks with increasing $q$. Nonetheless, our numerical analysis cannot completely rule out the possibility that the phase disappears at some large but finite $q$, rather than only in the strict XY limit.

\begin{figure}[htb]
    \centering
    \begin{overpic}[width=0.9\linewidth]{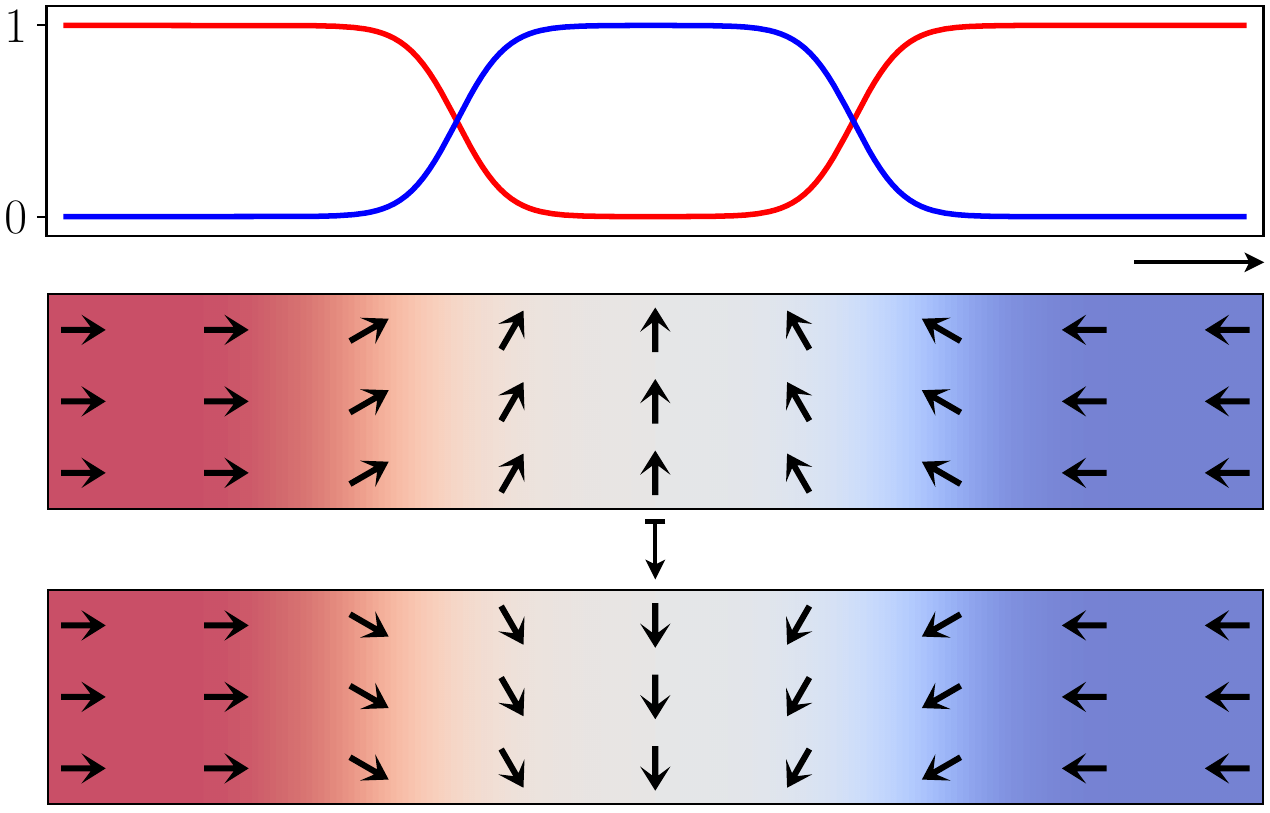}
    \put(82, 42.5){$x$}
    \put(83, 49){$\left| \sin(\frac{q \vartheta}{2})\right|$}
    \put(83, 57.8){$\left| \cos(\frac{q \vartheta}{2})\right|$}
    \put(54, 19.5){$\mathcal{T}$}
    \end{overpic}
    \caption{\textbf{Charged defects at the domain wall.} 
    In the $\mathbb{Z}_q^+$ AFM $\oplus$ $\mathbb{Z}_2$ NM phase, a smooth domain wall between the $\vartheta = 0$ and $\vartheta = \pi$ domains traps a $\mathcal{T}$ charge that flips sign under $\mathcal{T}$. Consequently, the $\mathbb{Z}_q^+$ order parameter $\cos(q\vartheta/2)$, which is finite in the bulk, vanishes at the domain wall, whereas the $\mathbb{Z}_q^-$ order parameter $\sin(q\vartheta/2)$, vanishing elsewhere, becomes non-zero on the domain wall.
    }
    \label{fig:domainwall}
\end{figure}

\subsection*{Enhanced symmetry and charged defects at the Landau-incompatible transitions}

The continuous transition between the Landau-incompatible $\mathbb{Z}_q^{\pm}$ AFM $\oplus$ $\mathbb{Z}_2$ NM phases exhibits key features of a deconfined quantum critical (DQC) transition. Most notably, as verified in the main text (see Fig.~\ref{fig:phase}(g)), it hosts an emergent $O(2)$ symmetry. This corresponds to the $O(2)$ symmetry of the critical Gaussian field theory. In the field-theoretic description (Eq.~\eqref{eq:S}), this can be understood as follows: the two Landau-incompatible ordered phases are characterized by the order parameters $\cos(q\vartheta/2)$ and $\sin(q\vartheta/2)$, respectively. The enhanced symmetry at the continuous transition between these two phases rotates between these two order parameters. This occurs when $\gamma=0$ (while $\cos(q\vartheta)$ remains relevant) and the additional perturbations $\cos(2q\vartheta)$ and $\cos(\phi)$ are irrelevant, resulting in a Gaussian field theory in the long-wavelength limit.

Another defining signature of a DQC transition is that topological defects of one Landau-incompatible order trap the charged degrees of freedom of the other, and their condensation drives the transition~\cite{SenthilDQCdoi:10.1126/science.1091806, SenthilDQCreview_doi:10.1142/9789811282386_0014}. Consequently, as one symmetry is restored, the other is simultaneously broken. As discussed in Ref.~\cite{prakash2024classicaloriginslandauincompatibletransitions}, this mechanism is indeed realized here.  Denoting $\vartheta = \arg(\tilde{m})$, 
where $\tilde{m}$ is defined in Eq.~\eqref{eq:afm_mag},
two of the vacua of the $\mathbb{Z}_q^+$ order, namely $\vartheta = 0$ and $\vartheta = \pi$, are invariant under the residual symmetry $\mathcal{T}: \vartheta \mapsto -\vartheta$. As a result, 
inside the $\mathbb{Z}_q^+$ AFM $\oplus$ $\mathbb{Z}_2$ NM phase,
any smooth domain wall between the $\vartheta = 0$ and $\vartheta = \pi$ regions traps a $\mathcal{T}$ charge (see Fig.~\ref{fig:domainwall}), and the $\mathbb{Z}_q^+$ order parameter $\cos(q\vartheta/2)$ that is finite elsewhere vanishes at the domain wall. The $\mathbb{Z}_q^-$ order parameter  $\sin(q\vartheta/2)$ has the opposite behavior and becomes finite only at the domain wall. Consequently, the continuous transition between the $\mathbb{Z}_q^{\pm}$ ordered phases, obtained by condensing these domain walls, simultaneously destroys the $\mathbb{Z}_q^+ $ order but produces the $\mathbb{Z}_q^-$ order.

\end{document}


\title{Supplemental material to ``Two-dimensional $J_1$-$J_2$ clock model: Enhanced symmetries, emergent orders, and Landau-incompatible transitions''}

\author{Pulloor Kuttanikkad Vishnu}
\affiliation{Department of Physics, Indian Institute of Technology Madras, Chennai 600036, India}
\affiliation{Center for Quantum Information, Communication and Computation (CQuICC), Indian Institute of Technology Madras, Chennai 600036, India}

\author{Abhishodh Prakash}
\affiliation{Harish-Chandra Research Institute, A CI of Homi Bhabha National Institute, Chhatnag Road, Jhunsi, Allahabad - 211019, India}

\author{Rajesh Narayanan}
\affiliation{Department of Physics, Indian Institute of Technology Madras, Chennai 600036, India}
\affiliation{Center for Quantum Information, Communication and Computation (CQuICC), Indian Institute of Technology Madras, Chennai 600036, India}

\author{Titas Chanda}
\email{titas.chanda@physics.iitm.ac.in}
\affiliation{Department of Physics, Indian Institute of Technology Madras, Chennai 600036, India}
\affiliation{Center for Quantum Information, Communication and Computation (CQuICC), Indian Institute of Technology Madras, Chennai 600036, India}

\begin{abstract}
In this supplemental material, we provide technical details on the classical Monte Carlo (CMC) simulations and the corner transfer matrix renormalization group (CTMRG) method. Additionally, we present supplementary results from CMC and CTMRG simulations, including the characterizations of the first-order phase transitions, to further support and complement the results discussed in the main text.
\end{abstract}

\date{\today}

\maketitle

\subsection{Deatils on the classical Monte Carlo simulations}

We employ classical Monte Carlo (CMC) simulations \cite{book_newman, metropolis_rosenbluth_jcp_53} to investigate different phases of the two-dimensional (2D) $J_1$-$J_2$ clock model, focusing on the $J_2/J_1 > 1/2$ regime. As discussed in the main text, frustration from competing interactions limits CMC updates to single-spin Metropolis flips, rendering the method inefficient for large system sizes, particularly in critical phases. Nonetheless, deep within each phase, CMC remains a valuable tool for the identification and characterization of different phases.

Here, we use $5 \times 10^5$ equilibration steps and $5 \times 10^4$ measurement steps for CMC, considering over 20 to 80 independent initial realizations. The system is initialized either in a random configuration at high temperatures or in one of the ordered ground-state ($T=0$) configurations, depending on the scenario. Temperature is then incrementally increased or decreased to compute various observables as functions of $T$.
Due to the limitations of single-spin flip algorithms, some configurations may get trapped in metastable states. We discard such configurations in our calculations.

We primarily compute two types of order parameters to identify phase transitions: one for detecting $\mathbb{Z}_q$ symmetry breaking and another for probing the breaking of sublattice $\mathbb{Z}_2$ symmetry, which governs the relative angle between the two $45^o$-rotated $\sqrt{2} \times \sqrt{2}$ sublattices. 

\begin{figure}[htb]
    \centering
    \includegraphics[width=\linewidth]{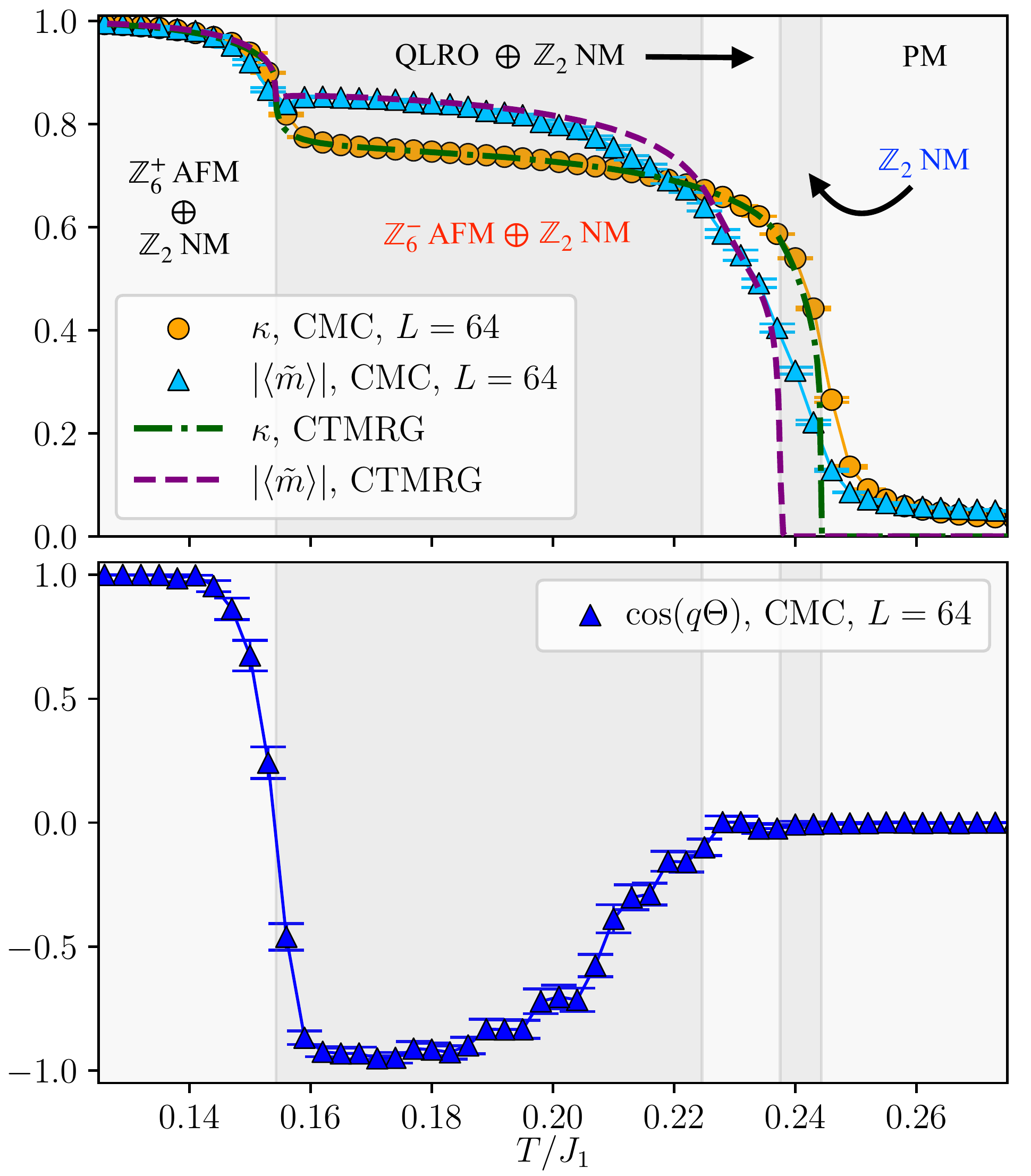}
    \caption{\textbf{Order parameters for $q=6$ at $J_2/J_1 = 0.52$.} We plot the order parameters $\kappa$ (top panel), $\cos(q \Theta)$ (bottom panel), and the absolute value $|\braket{\tilde{m}}|$ (top panel) for $q=6$, $J_2/J_1=0.52$, and system size $L=64$, as a function of $T/J_1$. Additionally, we include $\kappa$ and $|\braket{\tilde{m}}|$ computed using anisotropic CTMRG for comparison.}
    \label{fig:order_052}
\end{figure}

Since the system favors antiferromagnetic (AFM) ordering within each $\sqrt{2} \times \sqrt{2}$ sublattice for $J_2/J_1 > 1/2$ at low $T$, we compute the staggered magnetization within one such sublattice for each CMC snapshot:
\begin{equation} 
\tilde{m} = \frac{1}{L^2/2} \sum_{i_x, i_y} (-1)^{i_x+i_y} \exp(i \theta_{(i_x, i_y)}), 
\label{eq:afm_mag} 
\end{equation}
where $i_x$ and $i_y$ run over the rotated sublattice, and $L$ is the linear system size. We then compute the CMC average $\braket{\tilde{m}}$ over different snapshots. In the $\mathbb{Z}_q$-broken phase with an AFM order within the $\sqrt{2} \times \sqrt{2}$ sublattices, $\braket{\tilde{m}}$ must be different from zero.
In the main text, we have analyzed the phases of complex $\tilde{m}$ across different snapshots to construct the histograms. Here in the supplemental material, instead, we focus on the phase $\Theta$ of the CMC-averaged $\braket{\tilde{m}}$ and use $\cos(q \Theta)$ as an order parameter to signal $\mathbb{Z}_q$ symmetry breaking. To characterize the nematic (NM) order originating from spatial $\mathbb{Z}_2$ breaking we consider the order parameter as in the main text:
\begin{align}
    \kappa = \frac{1}{4} \langle & \cos\left(\theta_{\mathbf{i}} - \theta_{\mathbf{i} + \hat{x}}\right) + \cos\left(\theta_{\mathbf{i} + \hat{y}} - \theta_{\mathbf{i} + \hat{x} + \hat{y}}\right) \nonumber \\
    &- \cos\left(\theta_{\mathbf{i}} - \theta_{\mathbf{i} + \hat{y}}\right)  - \cos\left(\theta_{\mathbf{i} + \hat{x}} - \theta_{\mathbf{i} + \hat{x} + \hat{y}}\right) \rangle,
\end{align}
averaged over the entire lattice and CMC snapshots.

\begin{figure}
    \centering
    \includegraphics[width=\linewidth]{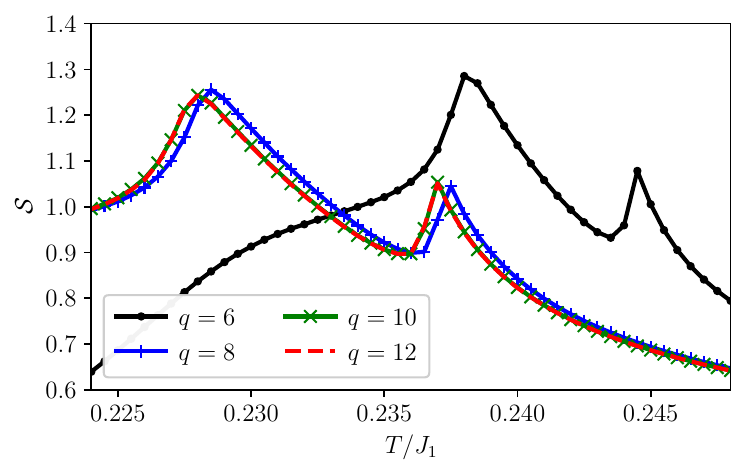}
    \caption{
    \textbf{The extent of the $\mathbb{Z}_2$ NM phase.} We plot the entanglement entropy $\mathcal{S}$ along the $J_2/J_1 = 0.52$ line for $q = 6, 8, 10,$ and $12$ as $T/J_1$ is varied from the QLRO $\oplus$ $\mathbb{Z}_2$ NM phase to the PM phase, passing through the intermediate $\mathbb{Z}_2$ NM phase. The region between the two peaks of each curve indicates the extent of the $\mathbb{Z}_2$ NM phase. Here, the anisotropic CTMRG method is employed on the original lattice with an MPS bond dimension of $\chi = 96$.
    }
    \label{fig:nm_phase}
\end{figure}

\begin{figure}[tb]
    \centering
    \includegraphics[width=\linewidth]{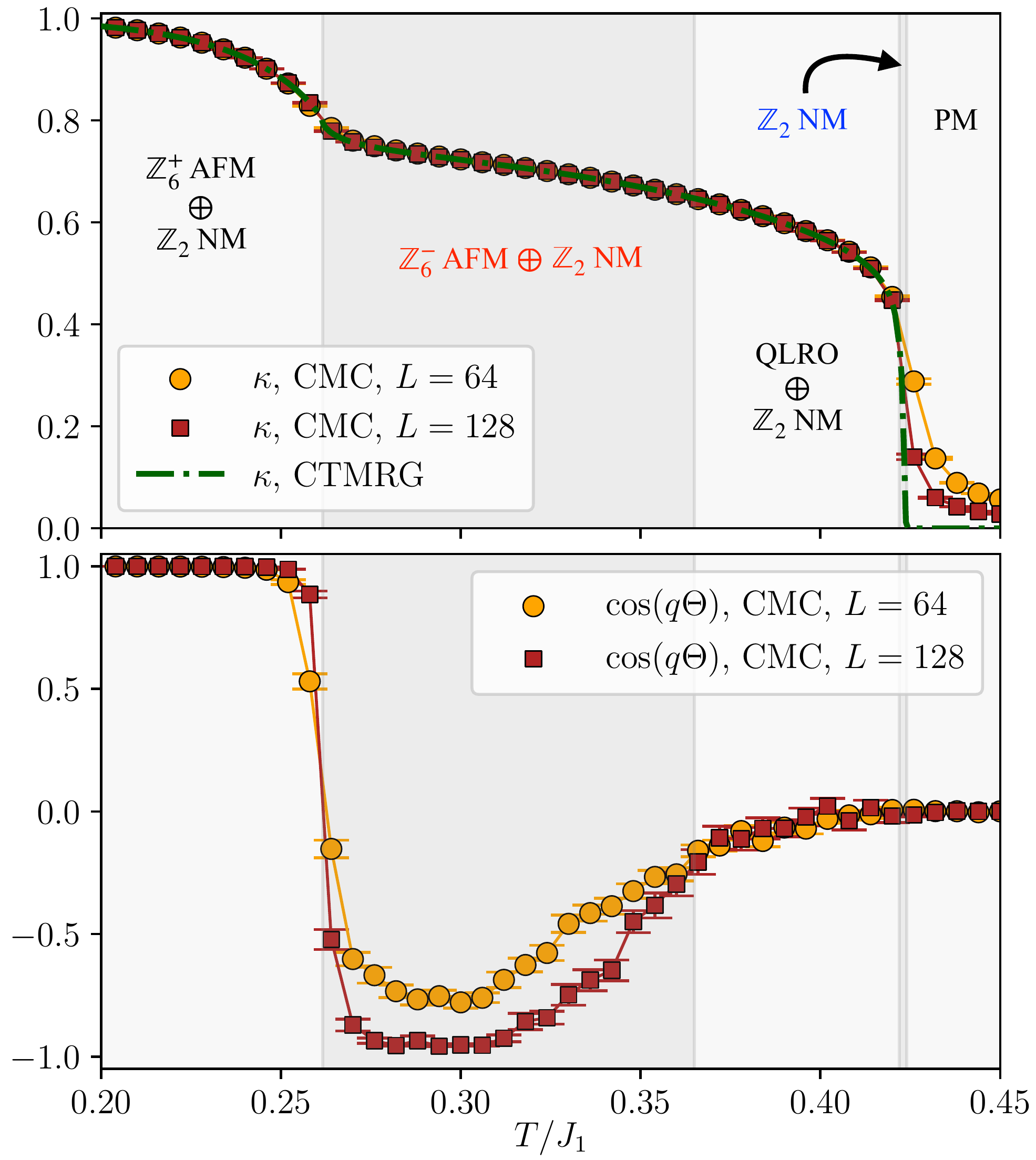}
    \caption{\textbf{Order parameters for $q=6$ at $J_2/J_1 = 0.6$.} Same as in Fig.~\ref{fig:order_06}, except here we consider $J_2/J_1 = 0.6$ and system sizes $L=64$ and $L=128$. Additionally, we omit $|\braket{\tilde{m}}|$ from the figure for better clarity.}
    \label{fig:order_06}
\end{figure}

In the low-$T$ $\mathbb{Z}_q^+$ AFM $\oplus$ $\mathbb{Z}_2$ NM phase, where effective clock angles are $2k\pi/q$ with $k = 0, 1, 2, \dots, q-1$ as in the microscopic model, $\cos(q \Theta)$ is expected to take values close to $+1$. Conversely, in the $\mathbb{Z}_q^-$ AFM $\oplus$ $\mathbb{Z}_2$ NM phase, where effective spins in the low-energy infrared limit are $(2k+1)\pi/q$, $\cos(q \Theta)$ should be close to $-1$. In the high-$T$ paramagnetic (PM) phase, where no long-range order persists, $\cos(q \Theta)$ should average to zero.

In Fig.~\ref{fig:order_052}, by fixing $q=6$, we present the order parameters $\kappa$ (top panel), $\cos(q \Theta)$ (bottom panel), and the absolute value $|\braket{\tilde{m}}|$ (top panel), computed via CMC for a linear system size of $L=64$, as $T/J_1$ is varied from the $\mathbb{Z}_q^+$ AFM $\oplus$ $\mathbb{Z}_2$ NM phase toward the high-$T$ PM phase. Additionally, we include $\kappa$ and $|\braket{\tilde{m}}|$ obtained using anisotropic corner transfer matrix renormalization group (CTMRG) with a $4 \times 4$ unit cell (see main text) for comparison.
The order parameters computed via CMC exhibit the expected behavior and show good agreement with the CTMRG results in the ordered $\mathbb{Z}_q^+$ AFM $\oplus$ $\mathbb{Z}_2$ NM and $\mathbb{Z}_q^-$ AFM $\oplus$ $\mathbb{Z}_2$ NM phases. However, due to the limitations of CMC, resolving the narrow regions of the critical quasi-long-range ordered (QLRO) phase with nematic order (QLRO $\oplus$ $\mathbb{Z}_2$ NM) and the ordered $\mathbb{Z}_2$ NM phase remains challenging. Therefore, we rely on CTMRG algorithms for the precise determination of phase boundaries and the identification of phases with narrow regions.
Our CTMRG calculations show that the extent of the $\mathbb{Z}_2$ NM phase for $q = 6$ is even smaller than that in the XY model. This is illustrated in Fig.~\ref{fig:nm_phase}, where we plot the entanglement entropy $\mathcal{S}$ along the $J_2/J_1 = 0.52$ line for $q = 6, 8, 10,$ and $12$. We observe that the extent of this phase remains essentially unchanged for $q = 10$ and $12$, suggesting that it coincides with that of the XY model. Moreover, for $q = 8$, both the location and the width of this phase are closer to the XY limit (i.e., to those for $q = 10$ or $12$) than in the case of $q = 6$. In the plot, the width of the $\mathbb{Z}_2$ NM phase for $q = 6$ is $\Delta (T/J_1) \approx 0.0065$, while for $q = 10$ or $12$ it is $\Delta (T/J_1) \approx 0.009$.

Figure~\ref{fig:order_06} closely resembles Fig.~\ref{fig:order_052}, except that it corresponds to $J_2/J_1 = 0.6$ and includes CMC results for two system sizes ($L=64, 128$). As the system size increases, the accuracy of the CMC results improves, and in the $\mathbb{Z}_q^-$ AFM $\oplus$ $\mathbb{Z}_2$ NM phase, $\cos(q \Theta)$ increasingly converges toward expected $-1$.

\subsection{Details on the corner transfer matrix renormalization group}

\begin{figure}[tb]
    \centering
    \includegraphics[width=\linewidth]{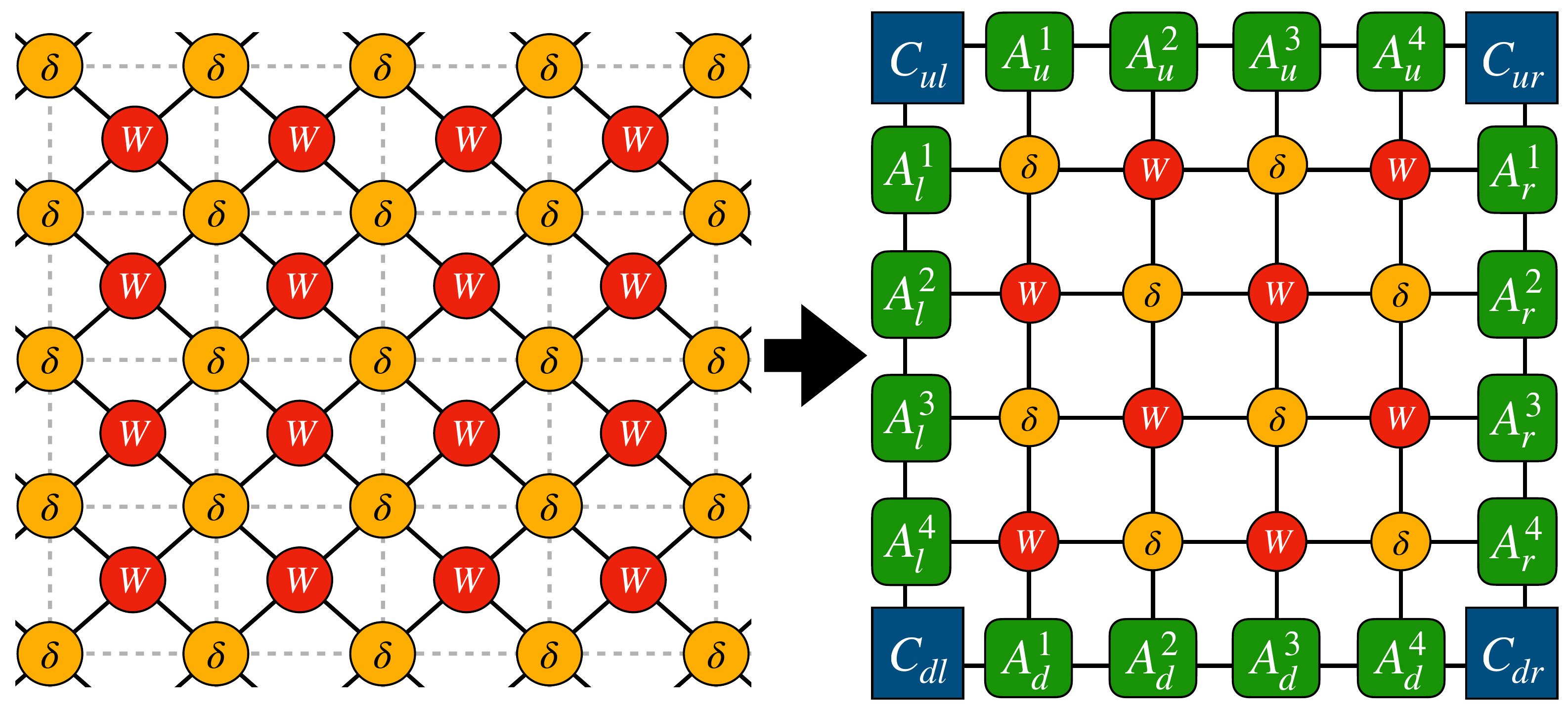}
    \caption{\textbf{TN representation of the $J_1$-$J_2$ clock model.}
    (Left panel) The partition function corresponding to the Hamiltonian~\eqref{eq:Hamil} can be written as a contraction of a 2D TN with two types of tensors, $W$ and $\delta$ (see text).
    (Right panel) The 2D TN can be contracted directly at the thermodynamic limit using directional anisotropic CTMRG, where we use $4 \times 4$ unit cell for accommodate the nematic order.
    }
    \label{fig:ani_partition}
\end{figure}

The partition function of the 2D $J_1$-$J_2$ $q$-state clock model (see main text):
\begin{equation} 
H = -J_1 \sum_{\braket{\mathbf{i}, \mathbf{j}}} \cos \left( \theta_{\mathbf{i}} - \theta_{\mathbf{j}} \right) + J_2 \sum_{\langle \braket{\mathbf{i}, \mathbf{j}} \rangle} \cos \left( \theta_{\mathbf{i}} - \theta_{\mathbf{j}} \right), 
\label{eq:Hamil} 
\end{equation}
with $\theta_{\mathbf{i}} = 2\pi k_{\mathbf{i}}/q$, can be expressed as a 2D tensor network (TN), as illustrated in the left panel of Fig.~\ref{fig:ani_partition}~\cite{li_pre_2021}. In Fig.~\ref{fig:ani_partition} (left panel), the tensor $W$ is defined as:
\begin{align} 
W_{ijkl} = \exp&\Bigg(\beta \frac{J_1}{2}\Big(\cos(\theta_i - \theta_j) + \cos(\theta_j - \theta_k) \nonumber \\ 
&+ \cos(\theta_k - \theta_l) + \cos(\theta_l - \theta_i)\Big) \nonumber \\ 
&- \beta J_2 \Big( \cos(\theta_i - \theta_k) + \cos(\theta_j - \theta_l)\Big) \Bigg), 
\end{align}
where $(i, j)$, $(j, k)$, $(k, l)$, and $(l, i)$ represent the tensor indices for nearest-neighbor (NN) spin pairs, while $(i, k)$ and $(j, l)$ correspond to next-nearest-neighbor (NNN) diagonal spin pairs. The $\delta$ tensor is defined as:
\begin{align}
    \delta_{ijkl} & = 1, \text{ if } i=j=k=l, \nonumber \\
                  & = 0, \text{ otherwise.}
\end{align}

This 2D TN, after a $45^{o}$ rotation, can be contracted efficiently using CTMRG~\cite{Nishino1996, Nishino1997, Corboz2014, Fishman2018} directly at the thermodynamic limit (Fig.~\ref{fig:ani_partition} (right panel)). Since the TN does not have $D_4$ symmetry (i.e., it is not isotropic), we consider the directional anisotropic CTMRG proposed by Corboz \textit{et. al.}~\cite{Corboz2014} and later improved in Ref.~\cite{Fishman2018}. Specifically, we use the algorithm described in Appendix A of Ref.~\cite{Fishman2018}.

To account for the spatial $\mathbb{Z}_2$-broken nematic or stripe order, we employ a $4 \times 4$ unit cell in the anisotropic CTMRG calculations, as depicted in the right panel of Fig.~\ref{fig:ani_partition}. In Fig.~\ref{fig:ani_partition} (right panel), the $(A_u, A_d, A_l, A_r)$'s represent the boundary fixed-point matrix-product state (MPS) tensors, while the $C$'s correspond to the corner transfer matrices (CTMs).
The increased unit-cell size makes the anisotropic CTMRG computationally demanding, and we therefore restrict the MPS bond dimension to $\chi = 128$ for $q=6$ for the results shown in the main text.

\begin{figure}
    \centering
    \includegraphics[width=0.75\linewidth]{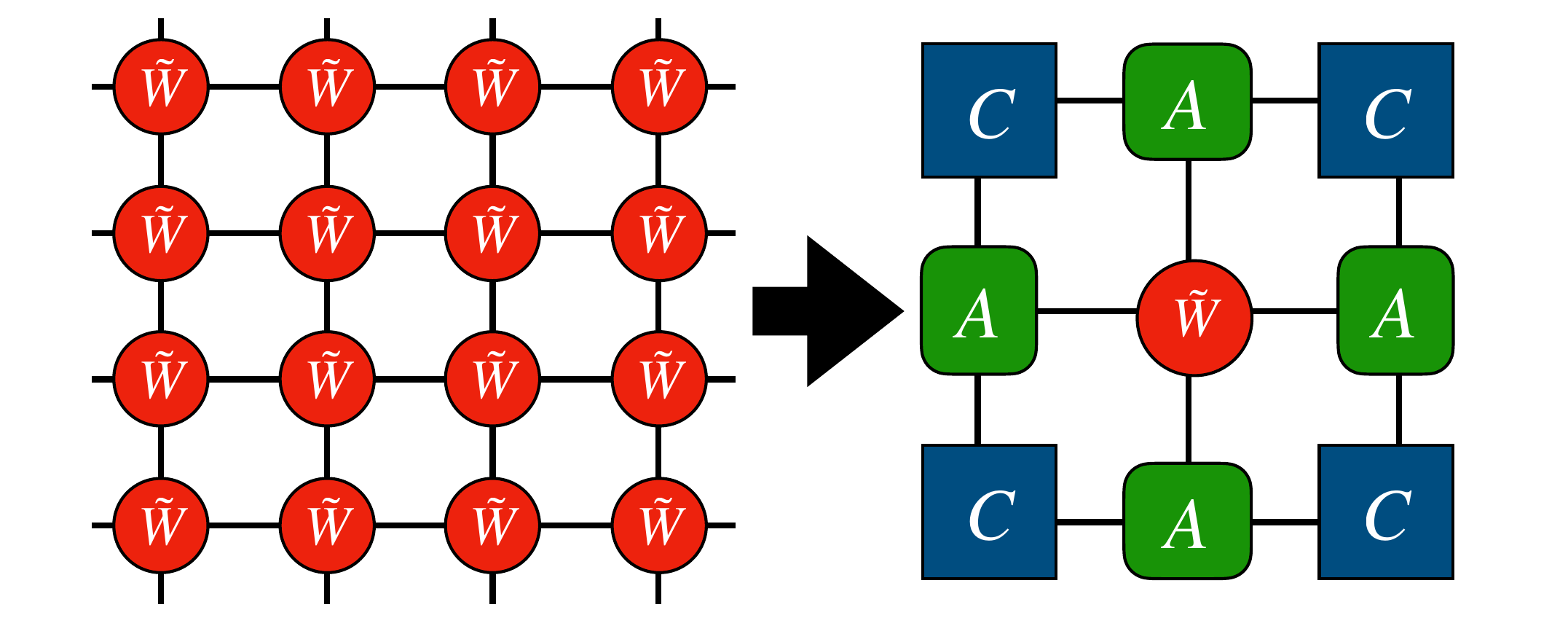}
    \caption{\textbf{TN representation of the dual system.} The partition function corresponding to the Hamiltonian~\eqref{eq:Hamil_dual} can be written as a contraction of a 2D TN involving a single isotropic tensor $\tilde{W}$ (see text).
    (Right panel) The 2D TN can be contracted directly at the thermodynamic limit using isotropic CTMRG.}
    \label{fig:dual_partition}
\end{figure}

Alternatively, we consider the dual system by defining bond degrees of freedom as $\sigma_{(\mathbf{i}, \hat{\delta})} := \left(\theta_{\mathbf{i}} - \theta_{\mathbf{i}+\hat{\delta}}\right)$ with $\hat{\delta} = \hat{x}, \hat{y}$~\cite{Chen2017, Li2020}. The corresponding dual Hamiltonian is given by:
\begin{align}
H_{D} = -J_1 \sum_{\mathbf{b}} \cos \sigma_{\mathbf{b}} + \frac{J_2}{2} \sum_{\langle \mathbf{b}, \mathbf{b}' \rangle} \cos\left(\sigma_{\mathbf{b}} + \sigma_{\mathbf{b}'} \right),
\label{eq:Hamil_dual} 
\end{align}
where the bond variables obey the local constraint:
\begin{equation}
\cos\left(\sigma_{(\mathbf{i}, \hat{x})} + \sigma_{(\mathbf{i}, \hat{y})} + \sigma_{(\mathbf{i} + \hat{x}, \hat{y})} + \sigma_{(\mathbf{i} + \hat{y}, \hat{x})} \right) = 1,
\label{eq:constraint}
\end{equation}
for each $\mathbf{i}$. The bond variables $\sigma_{\mathbf{b}}$ take values that are even multiples of $\pi/q$, as in the original formulation, with $\mathbf{b}$ denoting the bond index. The TN representation of the partition function for Eq.\eqref{eq:Hamil_dual} (see Fig.\ref{fig:dual_partition}, left panel) consists of a single rank-4 tensor $\tilde{W}$, defined as:
\begin{align}
\tilde{W}_{ijkl} = \exp\Bigg(&\beta \frac{J_1}{2} \Big (\cos\sigma_i + \cos\sigma_j + \cos\sigma_k + \cos\sigma_l\Big) \nonumber \\
&- \beta \frac{J_2}{2} \Big(\cos(\sigma_i + \sigma_j) + \cos(\sigma_j + \sigma_k) \nonumber \\
&+ \cos(\sigma_k + \sigma_l) + \cos(\sigma_l + \sigma_i)\Big)\Bigg),
\end{align}
if the constraint $\cos(\sigma_i + \sigma_j + \sigma_k + \sigma_l) = 1$ is satisfied; otherwise, $\tilde{W}_{ijkl} = 0$. Notably, this 2D TN, as depicted in the left panel of Fig.~\ref{fig:dual_partition}, exhibits full $D_4$ symmetry, allowing us to employ the fully isotropic original formulation of CTMRG~\cite{Nishino1996, Nishino1997}. In this approach, all CTMs and MPS tensors (see Fig.~\ref{fig:dual_partition}, right panel) are identical, enabling us to significantly increase the MPS bond dimension for enhanced numerical accuracy.

\begin{figure}
    \centering
    \includegraphics[width=0.5\linewidth]{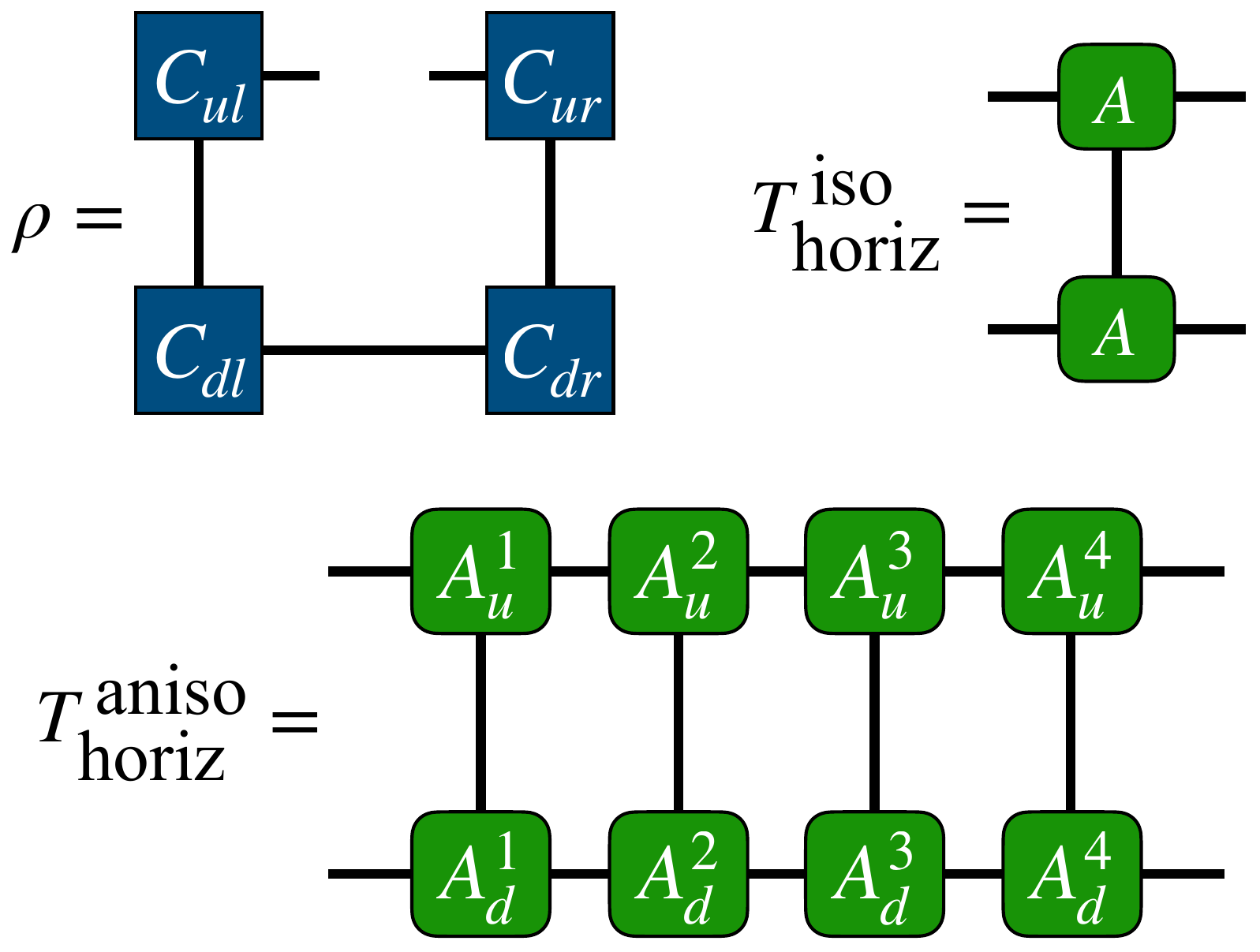}
    \caption{\textbf{Entanglement entropy and transfer matrix.}
    The reduced density matrix $\rho$ is obtained by contracting all CTMs in the manner shown in the figure. In the case of isotropic CTMRG, all CTMs are identical, simplifying the contraction process. The transfer matrix $T_{\text{horiz}}$ along the horizontal direction is constructed using the up and down MPS tensors. For anisotropic CTMRG, however, the full unit cell must be considered, requiring the contraction of all up and down MPS tensors, as illustrated in the figure. Similarly one can construct the transfer matrix long the vertical direction using left and right MPS tensors.}
    \label{fig:ee_t}
\end{figure}

In the CTMRG framework, free energy (per site), observables and correlation functions can be directly computed from the CTMs and MPS tensors; we refer to Refs.~\cite{Okunishi2022, Fishman2018, Xiang2023} for further details.
For the entanglement entropy (EE) of the boundary MPS, as reported in the main text, we obtain the reduced density matrix $\rho$ from the fixed-point CTMs (see Fig.~\ref{fig:ee_t}). The EE is then given by:
\begin{equation}
\mathcal{S} = - \sum_{i} \lambda_i \log \lambda_i,
\end{equation}
where $\lambda_i$ are the eigenvalues of $\rho$.
To compute the correlation length from the fixed-point boundary MPS, we first construct the transfer matrix $T$ (see Fig.~\ref{fig:ee_t}) in either the horizontal or vertical direction using the MPS tensors. The correlation length $\xi$ (along the corresponding direction) is then given by:
\begin{equation}
\xi = - \frac{1}{\log(|\epsilon_1|/|\epsilon_0|)},
\end{equation}
where $\epsilon_0$ is the dominant eigenvalue of $T$, and $\epsilon_1$ is the second-largest eigenvalue, ensuring that $|\epsilon_1| \neq |\epsilon_0|$.

\begin{figure}[t]
    \centering
    \includegraphics[width=\linewidth]{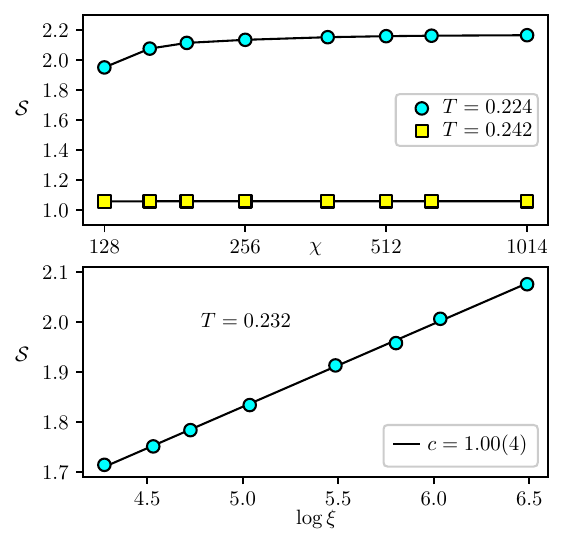}
    \caption{\textbf{Scaling of entanglement entropy.} We plot the EE at different phases for $J_2/J_1 = 0.52$ calculated for the dual system with isotropic CTMRG. 
    (Top panel) The saturation of EE with increasing $\chi$ confirms that the $\mathbb{Z}_q^-$ AFM $\oplus$ $\mathbb{Z}_2$ NM ($T=0.224$) and the $\mathbb{Z}_2$ NM ($T=0.242$) truly ordered non-critical phases. (Bottom panel) Critical EE scaling inside the QLRO $\oplus$ $\mathbb{Z}_2$ NM phase shows that the phase is described by a CFT with central charge $c=1$.
    }
    \label{fig:ee_scaling}
\end{figure}

In the top panel of Fig.~\ref{fig:ee_scaling}, we show the saturation of EE with increasing $\chi$ in the $\mathbb{Z}_q^-$ AFM $\oplus$ $\mathbb{Z}_2$ NM and the $\mathbb{Z}_2$ NM phases, confirming that these phases are non-critical. On the contrary, in QLRO $\oplus$ $\mathbb{Z}_2$ NM phase (or standard QLRO for $J_2/J_1 < 1/2$, not shown), the EE shows critical scaling~\cite{callan_geometric_1994, vidal_PRL_2003, calabrese_entanglement_2004}: 
\begin{equation}
    \mathcal{S} = \frac{c}{6} \log \xi + b,
\end{equation}
where $c$ is the central charge for the corresponding conformal field theory (CFT) with $b$ being a non-universal constant. By fitting this scaling law inside the QLRO $\oplus$ $\mathbb{Z}_2$ NM phase in the bottom panel of Fig.~\ref{fig:ee_scaling}, we verify that this phase is described by a CFT with $c=1$.

\begin{figure}[t]
    \centering
    \includegraphics[width=\linewidth]{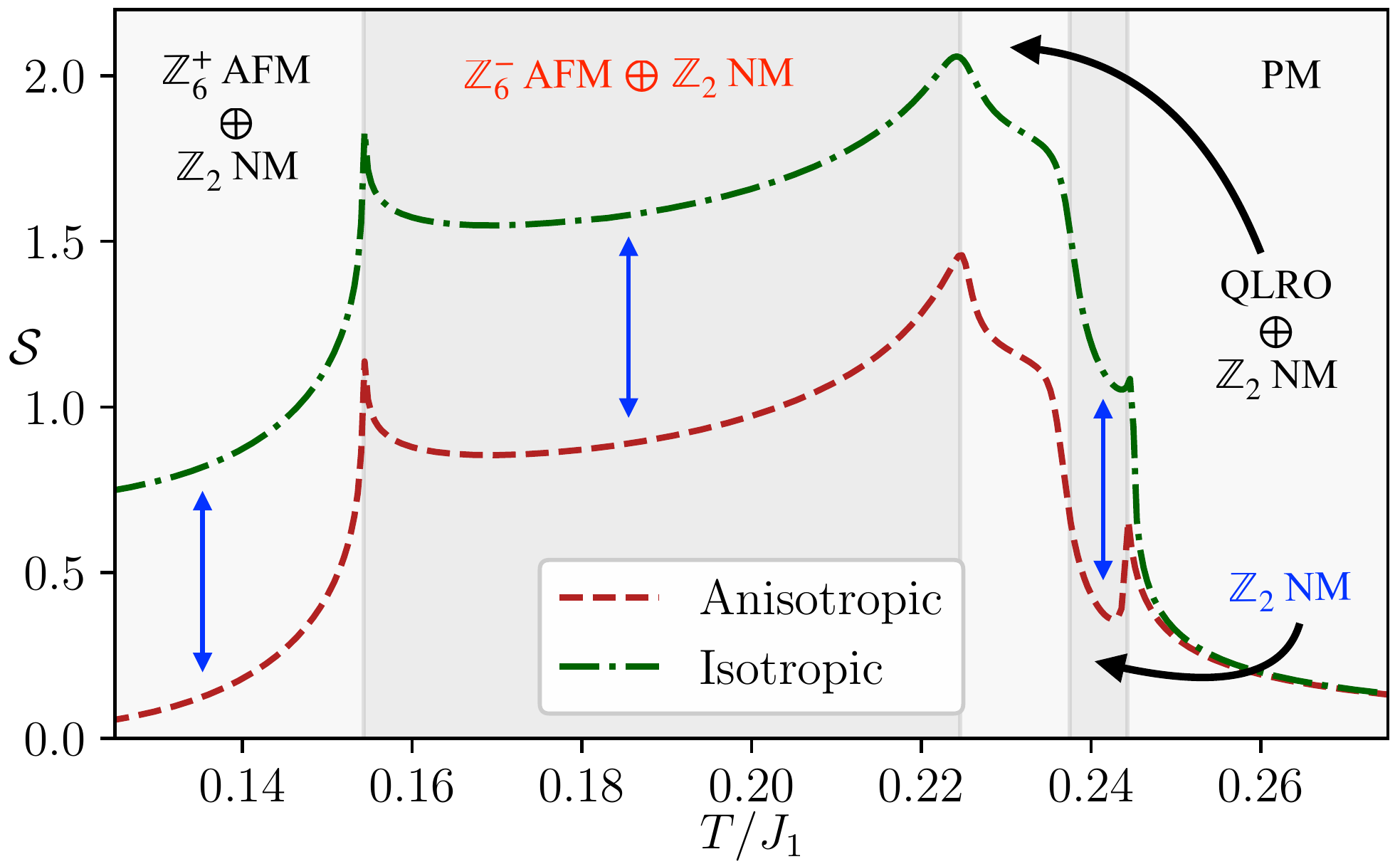}
    \caption{\textbf{Entanglement entropy in the dual system.}
    We plot the EE at $J_2/J_1 = 0.52$ in the dual system, computed using both anisotropic and isotropic CTMRG. In the non-critical nematic phases, their difference is $\log 2$ (indicated by blue arrows), whereas they coincide in the PM phase. Even in the critical QLRO $\oplus$ $\mathbb{Z}_2$ NM phase, the difference remains very close to $\log 2$.
    }
    \label{fig:ent_comp}
\end{figure}

Furthermore, we note that in the dual system, $\mathbb{Z}_q$ symmetry remains unbroken. However, the spatial $\mathbb{Z}_2$ symmetry between the sublattices can still break. When analyzed using isotropic CTMRG, this $\mathbb{Z}_2$ breaking results in an additional $\log 2$ contribution to the EE in the nematic phases, as the boundary MPS obtained from isotropic CTMRG is an equal superposition of the two ordered states.
On the other hand, in Fig.1(g) of the main text, we present the EE for the dual system computed using anisotropic CTMRG with a $2 \times 2$ unit cell. This minimal unit cell size is sufficient to break the spatial $\mathbb{Z}_2$ symmetry, leading to a $\log 2$ reduction in EE in the nematic phases compared to isotropic CTMRG. This behavior is explicitly demonstrated in Fig.~\ref{fig:ent_comp}.

\subsection{Characterizations of the first-order phase transitions}

\begin{figure}[htb]
    \centering
    \begin{overpic}[width=1\linewidth]{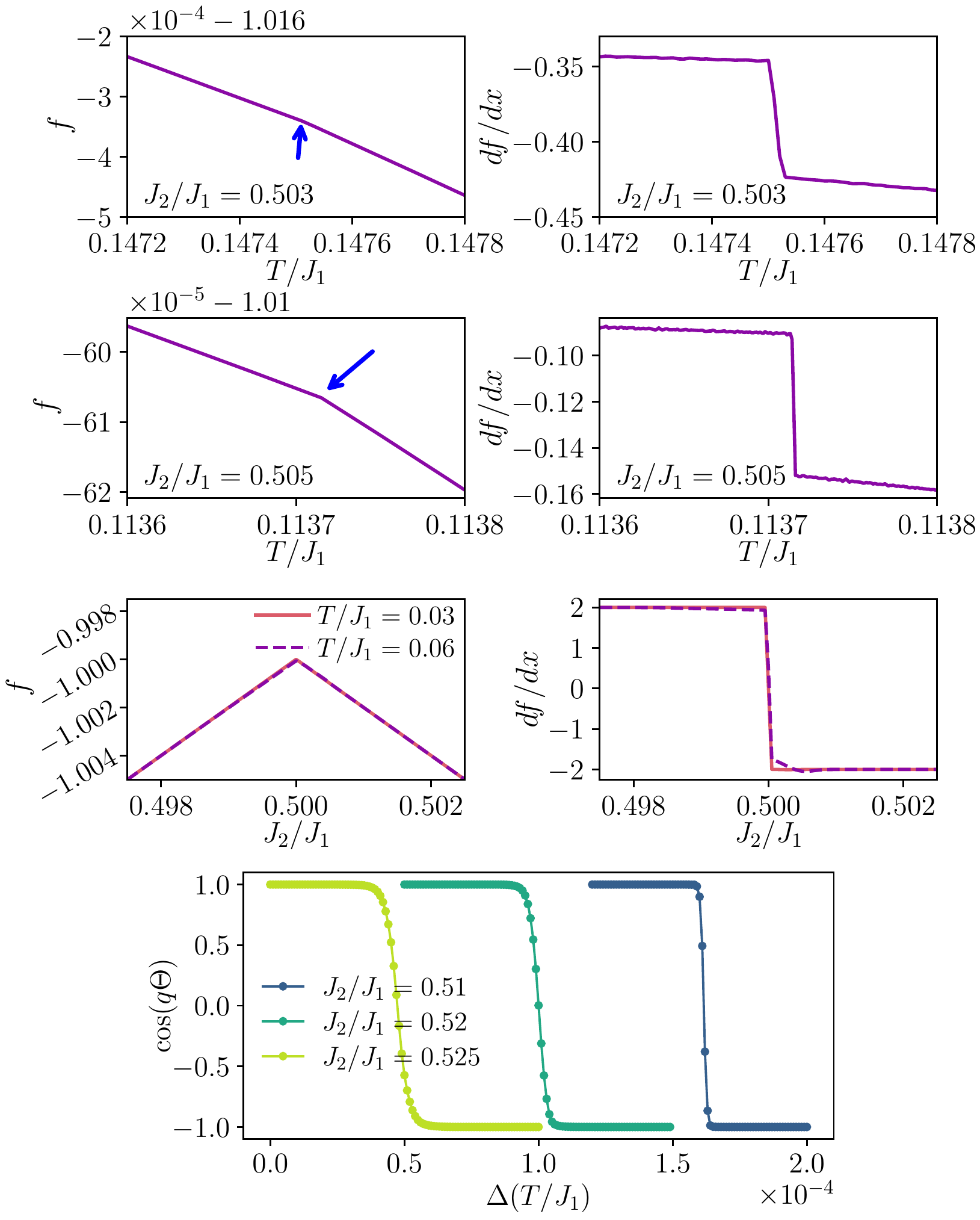}
    \put(33.5, 94){(a)}
    \put(72, 94){(d)}
    \put(33.5, 70.5){(b)}
    \put(72, 70.5){(e)}  
    \put(11, 47.5){(c)}           
    \put(72.3, 47.5){(f)}
    \put(64, 25){(g)}
    \end{overpic}
    \caption{\textbf{First order transitions.} 
    Free energy per site $f$, (a)-(c), and its first derivative $df/dx$, (d)-(f), across different first-order transitions, where $x$ denotes the parameter on the horizontal axis.
    The left column shows the behavior of (a) $f$  and (d) $df/dx$  at the transition between the $\mathbb{Z}_q^{-}$ AFM $\oplus$ $\mathbb{Z}_2$ NM and the $\mathbb{Z}_2$ NM phases. 
    The middle column, (b) and (e), displays the corresponding behavior for the transition between the $\mathbb{Z}_q^{\pm}$ AFM $\oplus$ $\mathbb{Z}_2$ NM phases at low $J_2/J_1 = 0.505$. 
    The right column illustrates (c) $f$  and (f) $df/dx$  across the low-$T$ transitions at $J_2/J_1 = 0.5$, specifically the $\mathbb{Z}_q^+$ FM ($T/J_1 = 0.03$) and QLRO ($T/J_1 = 0.06$) phases into the $\mathbb{Z}_q^+$ AFM $\oplus$ $\mathbb{Z}_2$ NM phase. 
    In the (a) and (b), the blue arrows indicate the non-analytic kink in $f$.
    (g) The behavior of $\cos(q \Theta)$, see text, across the $\mathbb{Z}_q^+$ AFM $\oplus$ $\mathbb{Z}_2$ NM $\leftrightarrow$ $\mathbb{Z}_q^-$ AFM $\oplus$ $\mathbb{Z}_2$ NM transitions for three different values of $J_2/J_1$.  For $J_2/J_1 = 0.51$, the horizontal axis is $T/J_1 = 0.13085 + \Delta(T/J_1)$; for $J_2/J_1 = 0.52$, it is $T/J_1 = 0.1542 + \Delta(T/J_1)$; and for $J_2/J_1 = 0.525$, it is $T/J_1 = 0.1636 + \Delta(T/J_1)$.
    All quantities are calculated using anisotropic CTMRG in the original lattice with $\chi = 128$.
    }
    \label{fig:first_order}
\end{figure}

In the main text, we have characterized the critical transitions present in the system.  Here, we analyze the first-order transitions mentioned there for $q=6$. 
In Fig.~\ref{fig:first_order}(a)-(f), we present the free energy per site $f$ and its first derivative across different first-order transitions. 
As an example, the transition between the $\mathbb{Z}_q^-$ AFM $\oplus$ $\mathbb{Z}_2$ NM and the $\mathbb{Z}_2$ NM phases (panels (a),(d)) can be clearly identified as first-order, since the first derivative of $f$ shows a sharp discontinuity. 
A similar first-order character is also visible at the $\mathbb{Z}_q^+$ AFM $\oplus$ $\mathbb{Z}_2$ NM $\leftrightarrow$ $\mathbb{Z}_q^-$ AFM $\oplus$ $\mathbb{Z}_2$ NM transition for small $J_2/J_1 > 0.5$ (panels (b),(c)), as well as at the low-$T$ transitions at $J_2/J_1 = 0.5$ (panels (c),(f)).

On the other hand, the $\mathbb{Z}_q^+$ AFM $\oplus$ $\mathbb{Z}_2$ NM $\leftrightarrow$ $\mathbb{Z}_q^-$ AFM $\oplus$ $\mathbb{Z}_2$ NM transition at larger $J_2/J_1$ requires a more careful analysis. Using a grid spacing of $10^{-6}$, we find that the transition for $J_2/J_1 \gtrsim 0.51$ is continuous, as evidenced both by the free energy (not shown) and by the behavior of the average effective spin angle $\Theta := \mathrm{arg}\,\langle e^{i\theta} \rangle$, where $\langle e^{i\theta} \rangle$ is computed at a representative CTMRG site. In Fig.~\ref{fig:first_order}(g), we plot $\cos(q\Theta)$ across the transition for three representative values of $J_2/J_1$: 0.51, 0.52, and 0.525. Within the $\mathbb{Z}_q^{\pm}$ AFM $\oplus$ $\mathbb{Z}_2$ NM phases, $\cos(q\Theta)$ takes values $\pm 1$, respectively. While $\cos(q\Theta)$ remains smooth across the transition for $J_2/J_1 = 0.52$ and $0.525$, it develops a distinctly sharper discontinuity at $J_2/J_1 = 0.51$. We recognize that an even finer resolution would be required to fully characterize the transition at $J_2/J_1 = 0.51$, but due to computational constraints we do not pursue this further here. Nonetheless, we can safely conclude that the direct transition between the $\mathbb{Z}_q^{\pm}$ AFM $\oplus$ $\mathbb{Z}_2$ NM phases is first-order for $J_2/J_1 \lesssim 0.51$, whereas for larger values it becomes a Landau-incompatible continuous transition.

\bibliography{suppl.bbl}